\pdfoutput=1
\documentclass[a4paper,fleqn,usenatbib]{mnras}

\usepackage{newtxtext,newtxmath}
\usepackage{bm}
\usepackage{tikz}
\usetikzlibrary{patterns}
\usetikzlibrary{positioning}
\usetikzlibrary{external}
\usepackage{blindtext}
\usepackage{mathdots}
\usepackage{enumitem}
\usepackage[ruled]{algorithm2e}

\usepackage{scalefnt}
\usepackage{pgfplots}
\usepackage{pgfplotstable}
\usepgfplotslibrary{fillbetween}
\usepackage{subfig}

\usepackage[T1]{fontenc}
\usepackage{ae,aecompl}

\usepackage{pgfplots}

\usepackage{tikz}
\usetikzlibrary{positioning}
\usetikzlibrary{shadows}

\usepackage{graphicx}	% Including figure files
\usepackage{amsmath}	% Advanced maths commands
\newcommand{\cb}{\text{cb}}
\newcommand{\bc}{\text{bc}}
\newcommand{\geom}{{}_2F_1}

\title[Higher-order neutrino ICs]{Higher-order initial conditions with massive neutrinos}

\author[Elbers et al.]{Willem Elbers$^{1}$, Carlos S. Frenk$^{1}$, Adrian Jenkins$^{1}$, Baojiu Li$^{1}$, and Silvia Pascoli$^{2,3,4,5}$\\
$^{1}$Institute for Computational Cosmology, Department of Physics, Durham University, South Road, Durham, DH1 3LE, UK\\
$^{2}$Institute for Particle Physics Phenomenology, Department of Physics, Durham University, South Road, Durham, DH1 3LE, UK\\
$^{3}$Dipartimento di Fisica e Astronomia, Universit\`a di Bologna, via Irnerio 46, 40126 Bologna, Italy\\
$^{4}$INFN, Sezione di Bologna, viale Berti Pichat 6/2, 40127 Bologna, Italy\\
$^{5}$Theoretical Physics Department, CERN, Geneva, Switzerland}

\date{Last updated xx; in original form xx}

\pubyear{2022}

\pubyear{2022}

\definecolor{lightgrey}{RGB}{230, 230, 230}

\begin{document}
\label{firstpage}
\pagerange{\pageref{firstpage}--\pageref{lastpage}}
\maketitle

\begin{abstract}
The discovery that neutrinos have mass has important consequences for cosmology. The main effect of massive neutrinos is to suppress the growth of cosmic structure on small scales. Such growth can be accurately modelled using cosmological $N$-body simulations, but doing so requires accurate initial conditions (ICs). There is a trade-off, especially with first-order ICs, between truncation errors for late starts and discreteness and relativistic errors for early starts. Errors can be minimized by starting simulations at late times using higher-order ICs. In this paper, we show that neutrino effects can be absorbed into scale-independent coefficients in higher-order Lagrangian perturbation theory (LPT). This clears the way for the use of higher-order ICs for massive neutrino simulations. We demonstrate that going to higher order substantially improves the accuracy of simulations. To match the sensitivity of surveys like DESI and Euclid, errors in the matter power spectrum should be well below $1\%$. However, we find that first-order Zel'dovich ICs lead to much larger errors, even when starting as early as $z=127$, exceeding $1\%$ at $z=0$ for $k>0.5\text{ Mpc}^{-1}$ for the power spectrum and $k>0.1\text{ Mpc}^{-1}$ for the equilateral bispectrum in our simulations. Ratios of power spectra with different neutrino masses are more robust than absolute statistics, but still depend on the choice of ICs. For all statistics considered, we obtain $1\%$ agreement between 2LPT and 3LPT at $z=0$.
\end{abstract}

\begin{keywords}
methods: numerical -- cosmology: theory -- large-scale structure of Universe -- dark matter -- physical data and processes: neutrinos\vspace{-1.3em}
\end{keywords}

%%%%%%%%%%%%%%%%%%%%%%%%%%%%%%%%%%%%%%%%%%%%%%%%%%

%%%%%%%%%%%%%%%%% BODY OF PAPER %%%%%%%%%%%%%%%%%%

% Hi there! How is your day going?

\section{Introduction}
The neutrino content of the Universe, $\Omega_\nu = \sum m_\nu / (93\text{ eV}\,h^2)$, becomes a powerful probe for cosmology once the implied neutrino masses are confronted with data from neutrino oscillations \citep{esteban20} and the kinematics of $\beta$-decay \citep{aker21}. A non-zero detection of $\Omega_\nu$ would be consequential for fundamental physics. It would confirm that a background of relic neutrinos survived until the epoch of structure formation, provide insight into the origin of neutrino mass, and constrain the search for dark matter and dark sectors. Oscillation experiments provide a lower bound of $\sum m_\nu>0.058$ eV, while cosmology provides upper bounds of $\sum m_\nu<0.15$ eV or better assuming $\Lambda$CDM \citep{palanque20,choudhury20,porredon21,valentino21}, with ongoing and future surveys promising significant further improvement. Planck and future cosmic microwave background experiments, together with large-scale structure surveys like DESI, Euclid, and Vera Rubin, could achieve sensitivities in the 0.01 - 0.02 eV range \citep{hamann12,abazajian15,brinckmann19,chudaykin19}. Such small shifts in neutrino mass correspond to tiny $0.5\%$ - $1.5\%$ effects on the power spectrum of matter fluctuations on $0.1\text{ Mpc}^{-1}$ to $1\text{ Mpc}^{-1}$ scales, requiring theoretical predictions that are at least as accurate.

With this goal in mind, many groups have studied the effects of massive neutrinos on large-scale structure. At early times and on large enough scales, perturbation theory is the method of choice for this purpose. Cosmological perturbation theory \citep{bernardeau02} is essential for providing analytical insight and a necessary complement to more expensive numerical simulations. The effects of neutrinos on the nonlinear matter power spectrum were first calculated at one-loop by \citet{saito08} and \citet{wong08}. Subsequent work has dealt more realistically with the neutrino phase-space distribution \citep{shoji10,dupuy14,blas14,fuhrer15,levi16,chen21}, which parallels similar efforts on the numerical simulations side. Other advances were made by including neutrinos in the effective field theory of large-scale structure \citep{senatore17,colas20} and using time renormalisation group perturbation theory \citep{lesgourgues09,upadhye19}, which improved agreement with $N$-body simulations. More closely related to this work, \citet{wright17} extended the hybrid COLA simulation method to cases with massive neutrinos using second-order Lagrangian perturbation theory (2LPT) and \citet{aviles20} incorporated nonlinear neutrino effects in Lagrangian perturbation theory up to third order (3LPT). On the numerical simulations side, where higher-order LPT has been used to great effect to produce accurate initial conditions (ICs) for conventional simulations without massive neutrinos \citep{scoccimarro98,sirko05,crocce06}, neutrino effects have not been included and higher-order LPT is therefore rarely used for neutrino simulations (but see \citealt{brandbyge08,yeche17}). In this work, we propose a novel scheme for generating $n$LPT ICs for neutrino simulations based on all-order recursive solutions in the small-scale limit. We also generate ICs based on a full calculation of scale-dependent neutrino effects in 2LPT, dealing with frame-lagging terms following \citet{aviles20}, and find near perfect agreement with our scheme in the final simulation product. This demonstrates that neutrino effects can be implemented beyond first order by working in the small-scale limit, paving the way for accurate neutrino simulation ICs.

$N$-body simulations are used to solve for the nonlinear gravitational dynamics of matter on small scales, where perturbation theory fails. Cosmological simulations with ICs based on LPT were pioneered by \citet{frenk83,klypin83} and \citet{efstathiou85}. Mixed dark matter simulations with sub-electronvolt neutrinos were first carried out by \citet{brandbyge08,brandbyge09,viel10}. We refer the reader to \citet{angulo21} for a review of neutrino simulation methods. As with perturbation theory, the accuracy of modern surveys places stringent demands on simulations, popularly expressed as a requirement for 1\% accurate calculations of the matter power spectrum \citep{schneider16}. A major source of uncertainty concerns the interface between perturbation theory and simulation, in the form of ICs, and associated transients \citep{scoccimarro98}. We may distinguish two major sources of uncertainty related to the choice of ICs \citep{efstathiou85,michaux21}. The first arise from discrepancies between the ICs and the actual nonlinear solution at the initial time. When the solution is calculated perturbatively at order $n$, this uncertainty can be understood as the truncation error introduced by neglecting terms of order $n+1$ and greater. The second source of uncertainty relates to discreteness effects that build up over time as the continuous fluid equations are solved by means of a discrete particle representation \citep{marcos06,garrison16}. There is a tension between these two, as early starts minimize truncation errors but entail larger discreteness errors, while late starts do the opposite. For example, the first-order solution of \citet{zeldovich70} has the largest possible truncation error, driving practitioners to start simulations early when higher-order corrections are small. However, such simulations manifest a greater dependence on particle resolution due to discreteness errors. While such errors can be corrected \citep{garrison16}, this reasoning provides strong motivation for using higher-order ICs at late times \citep{michaux21}.

Neutrinos complicate this picture in two ways. First, neutrinos introduce an additional length scale into the problem. Due to their large thermal velocities, neutrinos free stream out of potential wells, otherwise stated in terms of a suppression of clustering on scales smaller than a typical free-streaming length (\citealt{lesgourgues06}). This in turn causes a scale- and time-dependent suppression of dark matter and baryon clustering that must be accounted for in the initial conditions. \citet{zennaro17} showed how to incorporate such scale-dependence in a first order back-scaling procedure, but a consistent framework for higher-order ICs has thus far been lacking. We note that after we submitted our paper to the journal, \citet{heuschling22} presented a recipe for second-order neutrino ICs. Like us, they use a back-scaled transfer function for the cold dark matter and baryon species. The second complication is that late-time observables are more strongly correlated with the initial conditions and less determined by the internal structure of halos, when clustering is suppressed on small scales. This means that simulations with different neutrino masses are affected by errors to different degrees, contaminating ratios such as the suppression of the matter power spectrum. We will show that such ratios are more robust than absolute statistics, but still depend on the choice of initial conditions on small scales.

The paper is organized as follows. We begin by summarizing our recipe for generating higher-order ICs for neutrino simulations in section \ref{sec:ics}. The second part of the paper is concerned with a derivation of the higher-order solutions necessary for ICs, starting with the set-up in section \ref{sec:setup}, limiting solutions at all orders in section \ref{sec:approx}, and the full second-order solution in section \ref{sec:full}. The final third of the paper contains a systematic analysis of higher-order ICs in section \ref{sec:results}. Finally, we conclude in section \ref{sec:conclusion}. Throughout this paper, we use a default neutrino mass sum of $\sum m_\nu = 0.3$ eV to showcase our results, except where indicated otherwise.

\section{N-body Initial conditions}\label{sec:ics}

We begin by outlining our approach for setting up for 3-fluid ICs with cold dark matter (c), baryons (b), and neutrinos ($\nu$). Initially, we deal with a single cold fluid, described in terms of the the mass-weighted density contrast and velocity,
\begin{align}
    \delta_\cb = f_\text{c} \delta_\text{c} + f_\text{b} \delta_\text{b},\\
    \boldsymbol{v}_\cb = f_\text{c} \boldsymbol{v}_\text{c} + f_\text{b} \boldsymbol{v}_\text{b},
\end{align}

\noindent
where $f_\text{c} = \Omega_\text{c}/(\Omega_\text{c} + \Omega_\text{b})$ and $f_\text{b} = 1 - f_\text{c}$. In a final step, the cold fluid is separated into two components with distinct transfer functions. Our approach is based on a growing mode solution of the LPT equations in the small-scale limit, motivated by the hierarchy between the neutrino free-streaming scale and the nonlinear scale, $k_\text{fs} \ll k_\text{nl}$, at the redshifts relevant for ICs. In section \ref{sec:results}, we confirm that this is an excellent approximation suited for precision simulations. The recipe boils down to the following steps:
\begin{enumerate}[widest=99,itemindent=*,leftmargin=0pt]
    \item Compute a back-scaled transfer function $\delta_\cb(k)$
    \item Compute particle displacements via Eqs.~(\ref{eq:pot}--\ref{eq:cnfact_approx})
    \item Compute particle velocities via Eqs.~(\ref{eq:f_infty}--\ref{eq:veloc_corr})
    \item Perturb particle masses and velocities via Eqs.~(\ref{eq:component_c}--\ref{eq:veloc_diff})
\end{enumerate}

\noindent
These steps can be performed using a modified version of the \textsc{monofonIC} code \citep{michaux21}, which we have made publicly available\footnote{Up-to-date links to the software referenced in this paper are maintained at\\ \url{https://www.willemelbers.com/neutrino_ic_codes/}.}. We briefly discuss the steps in order and then deal with possible extensions in section \ref{sec:neutrinos} and \ref{sec:scales}.

\subsection{Transfer functions and back-scaling}\label{sec:TF}

In this paper, we follow the commonly used back-scaling approach. This approach begins by choosing a pivot redshift, typically $z=0$, where the simulation should reproduce linear theory on the largest scales. This is necessary because conventional $N$-body codes solve Newtonian equations and therefore fail to capture the large-scale general relativistic dynamics in which matter and radiation are coupled through the Einstein-Boltzmann equations. We note that there exist alternative solutions to this problem \citep{fidler17,brandbyge17,fidler19,tram19,partmann20} as well as fully relativistic $N$-body codes \citep{adamek17,barrera20,barrera20b}, which can avoid it altogether. In the back-scaling procedure, one uses a linear Einstein-Boltzmann code such as \textsc{class} \citep{lesgourgues11} or \textsc{camb} \citep{lewis11} to calculate the density transfer functions for each fluid species at $z^\text{pivot}$, which are then scaled back to the starting redshift of the simulation using the exact linear dynamics of the Newtonian code. For $\Lambda$CDM without neutrinos, this amounts to rescaling the dark matter transfer function by a constant growth factor ratio $D(z_i)/D(z^\text{pivot})$.

Adding massive neutrinos makes the linear solution scale-dependent, precluding a simple rescaling factor. Nevertheless, the same philosophy can be applied by solving the Newtonian dynamics of an $N$-body code with massive neutrinos at linear order. Following \citet{zennaro17}, we do this using a first-order Newtonian fluid approximation \citep{shoji10,blas14}, but see also \citet{heuschling22} for a relativistic formulation. This back-scaling method for neutrino cosmologies was first implemented in the \textsc{reps} code. To streamline the procedure for the end-user and to reduce the potential for human error, we built a lightweight back-scaling library \textsc{zwindstroom} that interfaces directly with \textsc{class} and the initial conditions generator \textsc{monofonIC}. The final result of these steps is a rescaled density transfer function $\delta_\cb(k)=D_\cb(k,z_i) / D_\cb(k,z^\text{pivot}) \cdot \delta_\cb(k,z^\text{pivot})$ for a cold dark matter-baryon fluid (cb), where the growth factor ratio is computed with \textsc{zwindstroom} and the transfer function with \textsc{class}.

\subsection{Displacements}\label{sec:disp}

The displacement field, $\bm{\psi}=\boldsymbol{x}-\boldsymbol{q}$, relates the particle position $\boldsymbol{x}$ to the corresponding Lagrangian coordinate $\boldsymbol{q}$. To determine $\bm{\psi}$, we first obtain the linear potential by solving
\begin{align}
    \nabla^2\varphi^{(1)}(\boldsymbol{q}) = \delta_\cb(\boldsymbol{q}). \label{eq:pot}
\end{align}

\noindent
Unless indicated otherwise, $\nabla=\nabla_{\boldsymbol{q}}$. We observe that $\varphi^{(1)}$ is not the gravitational potential, which also includes a neutrino contribution, but a notation that reflects the fact that we are solving for the displacements of cb fluid particles. Our fast approximate 3LPT \citep{buchert94,bouchet95,melott95} scheme for the displacement field in the presence of massive neutrinos has the simple form
\begin{align}
	\bm{\psi} &= \bm{\psi}^{(1)} + C_2 \bm{\psi}^{(2)} + C_3 \bm{\psi}^{(3a)} + C_2C^1_3 \bm{\psi}^{(3b)} + C_2\bm{\psi}^{(3c)}, \label{eq:3LPT}
\end{align}

\noindent
where $C_n$ are scale-independent factors that capture the absence of neutrino perturbations in the small-scale limit, $C^i_n=C_n/C_i$, and $\bm{\psi}^{(n)}$ have the same form in terms of $\varphi^{(1)}$ as in $\Lambda$CDM. In the notation of \citet{michaux21}, these are given by
\begin{alignat}{4}
    \bm{\psi}^{(1)}&=-\nabla\varphi^{(1)}\!\!, &    \bm{\psi}^{(2)} &= -\frac{3}{7}\nabla\varphi^{(2)},\;\; && &\\
    \bm{\psi}^{(3a)} &= \frac{1}{3}\nabla\varphi^{(3a)}\!\!,\,\, & \bm{\psi}^{(3b)} &= - \frac{10}{21}\nabla\varphi^{(3b)}\!\!,\,\, && \bm{\psi}^{(3c)} \,&= \frac{1}{7}\nabla\times\boldsymbol{A}^{(3)},
\end{alignat}

\noindent
with higher-order potentials given by
\begin{alignat}{3}
    &\nabla^2\varphi^{(2)} &&= \frac{1}{2}\left[\varphi^{(1)}_{,ii}\varphi^{(1)}_{,jj} - \varphi^{(1)}_{,ij}\varphi^{(1)}_{,ij}\right], \label{eq:approx_2lpt}\\
    &\nabla^2\varphi^{(3a)} &&= \text{det}\;\varphi^{(1)}_{,ij},\\
     &\nabla^2\varphi^{(3b)} &&= \frac{1}{2}\left[\varphi^{(2)}_{,ii}\varphi^{(1)}_{,jj} - \varphi^{(2)}_{,ij}\varphi^{(1)}_{,ij}\right],\\
      &\nabla^2\boldsymbol{A}^{(3)} &&= \nabla\varphi^{(2)}_{,i}\times\nabla\varphi^{(1)}_{,i}, \label{eq:approx_2lpt_3}
\end{alignat}

\noindent
where commas denote partial derivatives and we sum over repeated indices. In section \ref{sec:approx}, we show that $C_n$ can be expressed in terms of the neutrino fraction, $f_\nu=\Omega_\nu/\Omega_\text{m}$. The correction, as it turns out, is small and approximately linear in $f_\nu$:
\begin{align}
C_n &\cong 1 + \frac{2nf_\nu}{5(2n+3)}. \label{eq:cnfact_approx}
\end{align}

\noindent
For a minimal neutrino mass sum of $\sum m_\nu = 0.06\text{ eV}$, one finds $C_2-1=5\times10^{-4}$. For our fiducial mass sum of $\sum m_\nu=0.3\text{ eV}$, it is $0.3\%$. At $\sum m_\nu = 1\text{ eV}$, the effect is about one per cent. The third-order correction $C_3$ is larger, but since $\bm{\psi}^{(3)}$ is suppressed by another power of the growth factor, the overall impact is smaller.

\subsection{Velocities}

The velocity field is $\boldsymbol{v}_\cb=\mathrm{d}{\bm{\psi}}/\mathrm{d}t$. Given a satisfactory scheme for computing the displacement field, the time derivative can be evaluated numerically. This is our preferred method, since it requires no additional approximations. However, a faster method that avoids calculating higher order terms more than once is to use the asymptotic logarithmic growth rate
\begin{align}
    f_\infty=\lim_{k\to\infty}\frac{\mathrm{d}\log D_\cb(k,a)}{\mathrm{d}\log a}, \label{eq:f_infty}
\end{align}

\noindent
to convert displacements to velocities, setting
\begin{align}
    \boldsymbol{v}_\cb &= aHf_\infty\left[\bm{\psi}^{(1)} + 2C_2 \bm{\psi}^{(2)} \right. \label{eq:veloc}\\
    & \;\;\;\;\;\;\;\;\;\;\;\;\;\;\;\;\;\;\;\;\;\,\,  \left. +\,  3\left(C_3 \bm{\psi}^{(3a)} + C_2 C^1_3 \bm{\psi}^{(3b)} + C_2\bm{\psi}^{(3c)}\right)\right]. \nonumber
\end{align}

\noindent
By construction, this gives the correct particle velocities on small scales. To recover also the correct behaviour on horizon scales, we add a large-scale correction $\boldsymbol{v}_\cb^{(c)}$ given by
\begin{align}
\boldsymbol{v}_\cb^{(c)} &= aHf_\infty\nabla^{-2}\nabla(\theta_\cb - \delta_\cb), \label{eq:veloc_corr}
\end{align}

\noindent
where $\theta_\cb$ is the dimensionless energy flux transfer function computed with \textsc{class}. We verified that the resulting simulated power spectrum agrees with linear theory to better than $0.1\%$ at the pivot redshift of $z=0$ on large scales. However, this approximation neglects possible nonlinear effects of scale-dependent growth on particle velocities. Another alternative is to rescale the velocities by the scale-dependent growth rate \citep{zennaro17}, which faces a similar problem beyond linear order.

\subsection{Additional steps for 3-fluid ICs}\label{sec:three_fluid}

The steps above are sufficient for simulations with neutrinos and a single cold fluid. To separate this cold fluid into baryon and CDM components with distinct transfer functions, we follow the approach of \citet{hahn21}. In short, the component densities are related to the mass-weighted average via\footnote{We remind the reader that $f_\lambda = \Omega_\lambda/\Omega_\cb$ for $\lambda\in\{\text{c},\text{b}\}$ even as $f_\nu  = \Omega_\nu/ \Omega_\text{m} = \Omega_\nu /(\Omega_\cb+\Omega_\nu)=1-f_\cb$. Furthermore, $\delta_\bc \neq \delta_\cb$ and $\boldsymbol{v}_\bc\neq\boldsymbol{v}_\cb$.}
\begin{align}
    \delta_\text{c} &= \delta_\cb - f_\text{b}\delta_\bc, \label{eq:component_c}\\
    \delta_\text{b} &= \delta_\cb + f_\text{c}\delta_\bc,
\end{align}

\noindent
where the difference variable, $\delta_\bc = \delta_\text{b} - \delta_\text{c}$, is constant at first order. The velocity difference too is conserved and vanishes at all orders: $\boldsymbol{v}_\bc = \boldsymbol{v}_\text{b} - \boldsymbol{v}_\text{c} = 0$. These results, derived for $\Lambda$CDM without massive neutrinos \citep{rampf21}, carry over to the neutrino case, essentially due to the fact that the neutrino contribution cancels in the difference equations (Appendix \ref{sec:three_fluid_considerations}). The transfer function difference, $\delta_\bc(k)=\delta_\text{b}(k)-\delta_\text{c}(k)$, is computed with \textsc{class} at the pivot redshift and, since it is conserved, is not scaled back.

After assigning displacements and velocities to both particle species using the mass-weighted average fields, the density difference is implemented by setting the masses to
\begin{align}
    m_\lambda(\boldsymbol{q}) = \bar{m}_\lambda\big[1 + \delta_\lambda(\boldsymbol{q}) - \delta_\cb(\boldsymbol{q})\big], \label{eq:perturb_masses}
\end{align}

\noindent
with $\bar{m}_\lambda$ the mean particle mass for type $\lambda\in\{\text{c},\text{b}\}$. Perturbing the masses, rather than the displacements, was found by \citet{hahn21} to limit discreteness errors.

By construction, Newtonian simulations with initial conditions set up using the above procedure, reproduce the expected evolution of two cold fluids with a shared velocity field and a relative density contrast that is approximately conserved. However, like the large-scale velocity correction \eqref{eq:veloc_corr}, a further modification is needed to bring the dynamics back into agreement with \textsc{class} at first order:
\begin{align}
    m_\lambda(\boldsymbol{q}) &\to m_\lambda(\boldsymbol{q}) + 2\bar{m}_\lambda\left[\left(\frac{D_\infty(z^\text{pivot})}{D_\infty(z_i)}\right)^{1/2}-1\right]\Theta_\lambda(\boldsymbol{q}),\\
    \boldsymbol{v}_\lambda(\boldsymbol{q}) &\to \boldsymbol{v}_\lambda(\boldsymbol{q}) + aHf_\infty\left(\frac{D_\infty(z^\text{pivot})}{D_\infty(z_i)}\right)^{1/2}\nabla^{-2}\nabla\Theta_\lambda(\boldsymbol{q}), \label{eq:veloc_diff}
\end{align}

\noindent
where $D_\infty(z_i)$ is the small-scale growth factor at the starting redshift $z_i$ and $\Theta_\text{c} = -f_\text{b}\theta_\bc$ and $\Theta_\text{b} = f_\text{c}\theta_\bc$. The difference, $\theta_\bc(k)=\theta_\text{b}(k)-\theta_\text{c}(k)$, of the dimensionless energy flux transfer functions is computed with \textsc{class} at the pivot redshift.

\subsection{Neutrino particles}\label{sec:neutrinos}

Massive neutrinos can be included in $N$-body codes using a variety of methods. The most common approach is to solve for the neutrino perturbations self-consistently by including them as a separate $N$-body particle species \citep{brandbyge08,viel10}. Initial conditions are then also needed for these neutrino particles. Capturing the full neutrino phase-space distribution is non-trivial even in linear theory and it is therefore not sufficient to compute only the first two moments, as is done for baryons and CDM. Accurate neutrino particle initial conditions can be generated by integrating geodesics from high redshift \citep{ma94a,adamek17}, where the perturbed phase-space distribution can be expressed analytically \citep{ma95}, but care must be taken that the equations of motion remain valid in the ultra-relativistic r\'egime (Elbers, in prep.). This procedure can be carried out efficiently using our \textsc{FastDF} code. We stress that the focus of this paper is on dark matter and baryon ICs and the results apply regardless of whether the neutrino implementation uses particles.

\subsection{Scale-dependent effects}\label{sec:scales}

Finally, we verified the approximations above by performing a full calculation of scale-dependent effects on the second-order displacement field. This is done by replacing \eqref{eq:approx_2lpt} with a convolution of two copies of the first-order potential $\varphi^{(1)}(\boldsymbol{k})$, modulated by kernels $D^{(2)}_{A}(\boldsymbol{k}_1,\boldsymbol{k}_2)$ and $D^{(2)}_{B}(\boldsymbol{k}_1,\boldsymbol{k}_2)$, computed in section \ref{sec:full}. This numerical calculation is expensive, but we will show in section \ref{sec:results} that simulations with ICs based on the full calculation agree extremely well with those based on the approximate scheme described above. The reason for this is the hierarchy of scales, $k_\text{fs}\ll k_\text{nl}$, which implies that higher-order corrections are important only on scales where neutrinos do not cluster, at least at redshifts that are relevant for ICs. Since the overall impact of the third-order correction factor, $C_3$, is smaller than that of $C_2$ and given the excellent agreement between the full and approximate solutions at second order, we expect the difference to be even smaller at third order. At the same time, the triple convolutions required for the third-order solution would be prohibitively expensive and would require a different approach. For this reason, we only consider 2LPT in section \ref{sec:full}.

\section{Theoretical set-up}\label{sec:setup}

We now proceed with the set-up of a 3-fluid model, which is solved in section \ref{sec:lagrange}. We consider three fluids indexed by $\lambda\in\{\text{c},\text{b},\nu\}$ for cold dark matter, baryons, and neutrinos. Throughout, we will treat baryons like dark matter particles and denote the mass-weighted CDM-baryon fluid by subscript $\cb$. Let $\rho_\lambda(\boldsymbol{x})$ be the density, $\boldsymbol{u}_\lambda(\boldsymbol{x})$ the peculiar velocity flow, and $\bm{\sigma}_\lambda(\boldsymbol{x})$ the stress tensor. We also write $\delta_\lambda = \rho_\lambda/\bar{\rho}_\lambda-1$ for the density contrast.

\subsection{Euler equations}

Taking moments of the Boltzmann equation yields the Euler fluid equations \citep{bernardeau02}
\begin{align}
	&\partial_\tau\boldsymbol{u}_\lambda + \boldsymbol{u}_\lambda\cdot\nabla_{\boldsymbol{x}}\boldsymbol{u}_\lambda = -aH\boldsymbol{u}_\lambda - \nabla_{\boldsymbol{x}}\Phi - \frac{1}{\rho_\lambda}\nabla_{\boldsymbol{x}}(\rho_\lambda\bm{\sigma}_\lambda), \label{eq:og_euler}\\
	&\partial_\tau\delta_\lambda + \nabla_{\boldsymbol{x}}\cdot\left[(1+\delta_\lambda)\boldsymbol{u}_\lambda\right] = 0,  \;\;\;\;\;\;\;\;\;\text{for}\;\;\;\lambda\in\{\text{c},\text{b},\nu\}, \label{eq:og_cont}
\end{align}

\noindent
where $\tau$ is conformal time, $H=\partial_\tau a/a^2$ is the Hubble constant (given explicitly below) and $\Phi$ the Newtonian potential. While the neutrino distribution function and its higher-order moments are complicated, the stress tensor can be neglected for the cold dark matter and baryon fluids on the scales of interest, $\bm{\sigma}_\text{c}=\bm{\sigma}_\text{b}=0$. Taking the mass-weighted average of the cold dark matter and baryon equations, we obtain at all orders (see Appendix \ref{sec:three_fluid_considerations})
\begin{align}
	&\partial_\tau\boldsymbol{u}_\cb + \boldsymbol{u}_\cb\cdot\nabla_{\boldsymbol{x}}\boldsymbol{u}_\cb = -aH\boldsymbol{u}_\cb - \nabla_{\boldsymbol{x}}\Phi, \label{eq:euler1}\\
	&\partial_\tau\delta_\cb + \nabla_{\boldsymbol{x}}\cdot\left[(1+\delta_\cb)\boldsymbol{u}_\cb\right] = 0. \label{eq:euler2}
\end{align}

\noindent
The potential is given by Poisson's equation,
\begin{align}
	\nabla_{\boldsymbol{x}}^2\Phi(\boldsymbol{x}) = \frac{3}{2}\frac{\Omega_\text{m}H_0^2}{a}\delta_\text{m}(\boldsymbol{x}), \label{eq:og_poisson}
\end{align}

\noindent
in terms of the total matter density, $\delta_\text{m}=f_\cb\delta_\cb + f_\nu\delta_\nu$, which includes a massive neutrino contribution. To complete the system, we assume the linear response approximation for the neutrino density:
\begin{align}
	\delta_\nu(\boldsymbol{k}) &= \frac{\delta^\text{lin}_{\nu}(k)}{\delta^\text{lin}_\cb(k)}\delta_\cb(\boldsymbol{k}), \label{eq:linres_1}
\end{align}

\noindent
where $\delta^\text{lin}_\lambda(k)$ refers to the density transfer function of $\lambda\in\{\nu,\text{cb}\}$ computed in relativistic linear perturbation theory with \textsc{class}. The total matter density contrast is then
\begin{align}
	\delta_\text{m}(\boldsymbol{k}) &= \left[1 + \alpha(k)\right]f_\cb\delta_\cb(\boldsymbol{k}),  \label{eq:linres}
\end{align}

\noindent
where we have introduced the convenient notation $\alpha=f_\nu\delta^\text{lin}_{\nu}/(f_\cb\delta^\text{lin}_\cb)$ for the linear theory ratio. The linear response approximation is accurate while neutrinos and dark matter remain in phase, which is a reasonable assumption at the early times considered here (see below). Inserting this in \eqref{eq:og_poisson} yields
\begin{align}
	-k^2\Phi(\boldsymbol{k}) = \frac{B_0}{a}\left[1 + \alpha(k)\right]\delta_\cb(\boldsymbol{k}),  \label{eq:poisson}
\end{align}

\noindent
where $B_0=\frac{3}{2}(1-f_\nu)\Omega_\text{m}H_0^2$ is written in terms of present-day values. We look for a growing solution of the form $\delta_\cb(\boldsymbol{k},\tau) = D_\cb(k,\tau)\delta_\cb(\boldsymbol{k},\tau_0)$. Linearising (\ref{eq:euler1}-\ref{eq:og_poisson}), we find
\begin{align}
	\partial_\tau^2 D_\cb + aH\partial_\tau D_\cb = \frac{B_0}{a}(1+\alpha)D_\cb. \label{eq:growing}
\end{align}

\noindent
In contrast to the $\Lambda$CDM case, this equation is scale-dependent due to the appearance of $\alpha(k)$. To proceed, we will take the limit $k\to\infty$. Since $\lim_{k\to\infty}\alpha(k)=0$, we simply obtain
\begin{align}
	\partial_\tau^2 D_\infty + aH\partial_\tau D_\infty = \frac{B_0}{a}D_\infty \;\;\;\;\;\; (k\to\infty).  \label{eq:asymp_growing_tau}
\end{align}

\noindent
We denote the solution of \eqref{eq:asymp_growing_tau} by $D_\infty$ to indicate that this is the small-scale solution. At this point, an equally valid description could be given in the large-scale limit or indeed for an arbitrary pivot scale. We deliberately choose the small-scale limit for two reasons. First, most simulations are not large enough to realize the large-scale limit. Second, we are interested in nonlinear corrections to the initial conditions which are negligible on large scales.

\subsection{Asymptotic form}

We can find an analytic\footnote{A function $f$ is analytic at $x$ if the Taylor series of $f$ around $x$ converges to $f$ in a neighbourhood of $x$.} solution to \eqref{eq:asymp_growing_tau} if the contribution of radiation to the Hubble rate is neglected. We will return to this point further below. For now, let us assume that
\begin{align}
	H^2= H_0^2\left[\Omega_\Lambda + \frac{\Omega_\cb + \Omega_\nu}{a^3}\right]. \label{eq:Hubble}
\end{align}

\noindent
In this case, the growing mode can be expressed in terms of the hypergeometric function as (see Appendix \ref{sec:algebra})
\begin{align}
    D_\infty(a)=a^p\sqrt{1+\Lambda a^3}\geom\left(\frac{2p+7}{6}, \frac{2p+3}{6}, \frac{4p+7}{6},-\Lambda a^3\right)\!,
\end{align}

\noindent
with $\Lambda=\Omega_\Lambda/\Omega_\text{m}$ and $p=\sqrt{1+24(1-f_\nu)}/4-1/4$. This is normalized such that $\lim_{a\to0}D_\infty/a^p=1$. Taking $f_\nu=0$, we recover the $\Lambda$CDM solution with $p=1$ \citep{rampf15}. Taking instead $\Lambda\to0$, we recover the solution during matter domination (MD)
\begin{align}
	D_\infty(a) = a^p = a^{\sqrt{1+24(1-f_\nu)}/4-1/4}, \label{eq:D_first_order}
\end{align}

\noindent
which agrees with \citet{bond80}.

For $\Lambda$CDM without massive neutrinos, accurate nonlinear predictions can be made by substituting the growth factor for the scale factor, $a\to D$, in solutions obtained for the Einstein-de Sitter model. This is facilitated by using the growth factor as time variable (e.g. \citealt{matsubara15,rampf15,rampf21}). Here, we will pursue a similar strategy and make a change of time variables to $D_\infty$. Defining the quantity
\begin{align}
    g_\infty = \frac{2}{3}\frac{B_0}{a}\left(\frac{D_\infty}{\partial_\tau D_\infty}\right)^2 \label{eq:g_inf_def}
\end{align}

\noindent
and the new velocity variable $\boldsymbol{v}_\cb = \partial_{D_\infty}\boldsymbol{x}$, the fluid equations can be rewritten as
\begin{align}
	&\partial_{D_\infty}\boldsymbol{v}_\cb + \boldsymbol{v}_\cb\cdot\nabla_{\boldsymbol{x}}\boldsymbol{v}_\cb = -\frac{3g_\infty}{2D_\infty}(\boldsymbol{v}_\cb + \nabla_{\boldsymbol{x}}\varphi),  \label{eq:euler_Dinf_euler}\\
	&\partial_{D_\infty}\delta_\cb + \nabla_{\boldsymbol{x}}\cdot\left[(1+\delta_\cb)\boldsymbol{v}_\cb\right] = 0,  \label{eq:euler_Dinf_continuity}\\
	&\nabla_{\boldsymbol{x}}^2\varphi = \frac{\delta_\cb}{D_\infty}*(1+\alpha), \label{eq:euler_Dinf_poisson}
\end{align}

\noindent
where the rescaled potential $\varphi = a\Phi/(B_0 D_\infty)$ is given in terms of a convolution, denoted by $*$, of $\delta_\cb$ and the linear response $(1+\alpha)$. Although written in terms of $D_\infty$, this is completely general. 

\begin{figure}
	\normalsize
	\centering
	\includegraphics{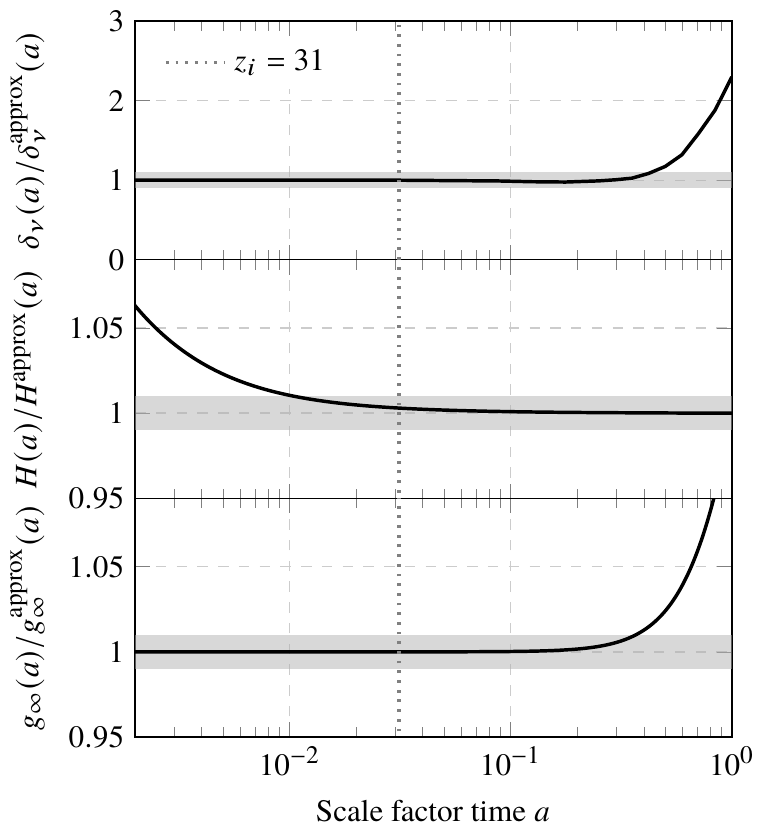}
	\caption{Accuracy of the linear response approximation \eqref{eq:linres_1} evaluated at $k=0.60\text{ Mpc}^{-1}$, compared to a reference simulation (top), of neglecting radiation in \eqref{eq:Hubble} for the Hubble rate (middle), and of \eqref{eq:early_g_infty} for the constant matter-dominated value for $g_\infty$. The vertical dotted line indicates the fiducial starting redshift of $z_i=31$. The neutrino mass sum is $\sum m_\nu = 0.3$ eV and the shaded region is $10\%$ (top) and $1\%$ (middle \& bottom).}
	\label{fig:approx_validity_1}
\end{figure}

Given suitable boundary conditions, Eqs.~(\ref{eq:euler_Dinf_euler}-\ref{eq:euler_Dinf_poisson}) are analytic at $D_\infty=0$. In particular, we require that $\delta^\text{ini}_\text{m}=\delta^\text{ini}_\cb=0$. This agrees with our use of growing mode solutions for particle displacements, $\boldsymbol{q}\mapsto\boldsymbol{q}+\bm{\psi}$, where the unperturbed particle grid represents a uniform density field. The scaling, $H^2\propto a^{-3}$, of the Hubble rate at early times ensures that such mass transport problems are well-posed \citep{brenier03,rampf15}. This scaling does not hold in the presence of radiation, a problem that already occurs in $\Lambda$CDM on account of the cosmic microwave background radiation, but is certainly made worse by the inclusion of massive neutrinos, which scale like radiation in the relativistic r\'egime. Therefore, we need to start the integration at a time when the relativistic contribution of neutrinos to the Hubble expansion can be neglected. Note that we make this assumption to ensure a consistent mathematical framework for the higher-order LPT solutions. However, it is not needed for the linear transfer functions, the back-scaling procedure or in the $N$-body code itself. In each of those cases, we do take the relativistic neutrino contribution into account.

Before proceeding, let us give the following convenient expression for $g_\infty$ in the limit $\Lambda\to0$:
\begin{align}
	g_\infty^{-1/2} = \frac{a^{3/2}H}{\sqrt{\tfrac{2}{3}B_0}}\frac{\mathrm{d}\log D_\infty}{\mathrm{d}\log a} = \frac{1}{4}\frac{\sqrt{1+24(1-f_\nu)}-1}{\sqrt{1-f_\nu}}. \label{eq:early_g_infty}
\end{align}

\noindent
Both numerator and denominator scale approximately as $(1-f_\nu)^{1/2}$. The numerator is simply the exponent of the growing mode in \eqref{eq:D_first_order}, while the dependence of the denominator can be traced to the appearance of $B_0$ on the right-hand side of \eqref{eq:asymp_growing_tau}. The resulting smallness of $g_\infty-1$ explains why neutrino corrections at $n$th order are small relative to $D_\infty^n$: the lack of neutrino clustering is largely compensated by slower growth of the linear solution. In the next section, we will validate the assumptions made up to this point.

\subsection{Validity of assumptions}

Central to the approach of section \ref{sec:lagrange} is the linear response approximation \eqref{eq:linres_1} for the nonlinear neutrino density, $\delta_\nu(\boldsymbol{k})$. This approximation is very accurate at early times, but underestimates neutrino clustering on small scales and neglects the phase shift between neutrinos and dark matter that builds up at late times (see Fig.~6 in \citealt{elbers20}). The top panel of Fig.~\ref{fig:approx_validity_1} shows the nonlinear neutrino density contrast, computed from a simulation with neutrino particles, relative to the linear neutrino response evaluated at $k=0.60\text{ Mpc}^{-1}$. The neutrino mass is $\sum m_\nu=0.3$ eV. The figure suggests that the approximation is valid at this scale up to $z\approx1.5$, when perturbation theory has presumably already broken down. Hence, approximation \eqref{eq:linres_1} is well-suited for our application at much higher redshifts.

A second approximation is that we neglect the contribution of the relativistic tail of the neutrino distribution to the Hubble rate in \eqref{eq:Hubble}. We reiterate that this approximation is only made for the calculation of the higher-order kernels and not in any of the calculations at first order. The middle panel of Fig.~\ref{fig:approx_validity_1} shows that this approximation is accurate to better than $1\%$ for $a>0.01$, for our default neutrino mass of $\sum m_\nu = 0.3$ eV. In particular, at the fiducial starting redshift of $z_i=31$, the error is $0.3\%$. We are helped in this regard by our preference for late starts.

Finally, we assume that $g_\infty$ is constant in section \ref{sec:approx}. The bottom panel of Fig.~\ref{fig:approx_validity_1} shows that this is an excellent approximation, except at late times during $\Lambda$-domination. The figure suggests that there is a window where all assumptions are valid, potentially allowing us to push to even later starts, with the breakdown of LPT likely being the limiting factor.

\section{Lagrangian approach}\label{sec:lagrange}

In the Lagrangian approach to gravitational instability \citep{zeldovich70,buchert89,moutarde91,bouchet92,gramann93,buchert93,bouchet95}, the objective is to describe fluid particle trajectories
\begin{align}
\boldsymbol{x}(\boldsymbol{q}) = \boldsymbol{q} + \bm{\psi}(\boldsymbol{q}),
\end{align}

\noindent
in terms of a displacement field $\bm{\psi}$. We use the Helmholtz decomposition, writing the Laplacian of a smooth vector field as
\begin{align}
    \nabla^2\bm{\psi} = \nabla\left(\nabla\cdot\bm{\psi}\right) - \nabla\times\left(\nabla\times\bm{\psi}\right).
\end{align}

\noindent
What remains is to solve for the longitudinal and transverse derivatives. The displacement is related to the Eulerian density, $\delta_\cb$, through the mass conservation equation
\begin{align}
	\delta_\cb(\boldsymbol{x}) = \frac{1}{J(\boldsymbol{q})} - 1, \label{eq:MA}
\end{align}

\noindent
where $J(\boldsymbol{q})$ is the determinant of the Jacobian of the coordinate transformation, $J_{ij}=\partial x_i/\partial q_j$, given by
\begin{align}
J = \mathrm{det}\;J_{ij} &= 1 + \psi_{i,i} + \frac{1}{2}\left[\psi_{i,i}\psi_{j,j}-\psi_{i,j}\psi_{j,i}\right] + \mathrm{det}\;\psi_{i,j}. \label{eq:jacobian}
\end{align}

\noindent
Let $(\partial/\partial D_\infty)_\text{L} = \left(\partial_{D_\infty} + \boldsymbol{v}_\cb\cdot\nabla_{\boldsymbol{x}}\right)$ be the Lagrangian derivative. The Lagrangian form of the Euler equation \eqref{eq:euler_Dinf_euler} can be written as
\begin{align}
    \mathcal{D}_\infty\boldsymbol{x} = -\frac{3g_\infty}{2D_\infty}\nabla_{\boldsymbol{x}}\varphi, \label{eq:lagrangian_acceleration}
\end{align}

\noindent
where we used $\boldsymbol{v}_\cb = (\partial \boldsymbol{x}/\partial D_\infty)_\text{L}$ and introduced the linear operator
\begin{align}
	\mathcal{D}_\infty = \left(\frac{\partial}{\partial D_\infty}\right)_\text{L}^2 + \frac{3g_\infty}{2D_\infty}\left(\frac{\partial}{\partial D_\infty}\right)_\text{L}. \label{eq:operator}
\end{align}

\noindent
Using \eqref{eq:euler_Dinf_poisson} and taking the divergence and curl of \eqref{eq:lagrangian_acceleration}, we find that the evolution of the displacement is governed by
\begin{align}
	\nabla_{\boldsymbol{x}}\cdot\mathcal{D}_\infty \boldsymbol{x}(\boldsymbol{q}) &= -\frac{3g_\infty}{2D_\infty^2}\left[\delta_\cb*(1+\alpha)\right](\boldsymbol{x}), \label{eq:semi_lagrange}\\
    \nabla_{\boldsymbol{x}}\times\mathcal{D}_\infty \boldsymbol{x}(\boldsymbol{q}) &= 0. \label{eq:semi_lagrange_trans}
\end{align}

\noindent
To facilitate a fully Lagrangian description, we define the frame-lagging terms \citep{aviles17,wright17}
\begin{align}
	F(\boldsymbol{q}) \equiv \left[\left(1/J-1\right)*\alpha\right](\boldsymbol{q}) - [\delta_\cb*\alpha](\boldsymbol{x}).
\end{align}

\noindent
Frame-lagging terms arise from mapping the Eulerian neutrino response to Lagrangian coordinates. We give explicit expressions up to second order in Appendix \ref{sec:FL}. Transforming the derivatives on the left-hand side of \eqref{eq:semi_lagrange} and \eqref{eq:semi_lagrange_trans} using $\partial_{x_i} = (\partial{q_j}/\partial{x_i})\partial_{q_j} = J^{-1}_{ij}\partial_{q_j}$ and using the Monge-Amp\`ere equation \eqref{eq:MA}, we write these equations in Lagrangian coordinates as
\begin{align}
	J^{-1}_{ij}\mathcal{D}_\infty \psi_{i,j} &= \frac{3g_\infty}{2D_\infty^2}\left[(1-1/J)*(1+\alpha) + F\right], \label{eq:compact_long}\\
    \epsilon_{ijk} J^{-1}_{jl}\mathcal{D}_\infty \psi_{k,l} &= 0. \label{eq:compact_trans}
\end{align}

\noindent
It will be the task in the following sections to find perturbative solutions for $\bm{\psi}$. We perform an expansion in displacements, writing
\begin{align}
    \bm{\psi} = \sum_{n=1}^\infty \bm{\psi}^{(n)}, \label{eq:powers}
\end{align}

\noindent
where $\bm{\psi}^{(n)}$ is of order $\big[\bm{\psi}^{(1)}\big]^n$.

\begin{figure}
	\normalsize
	\centering
	\includegraphics{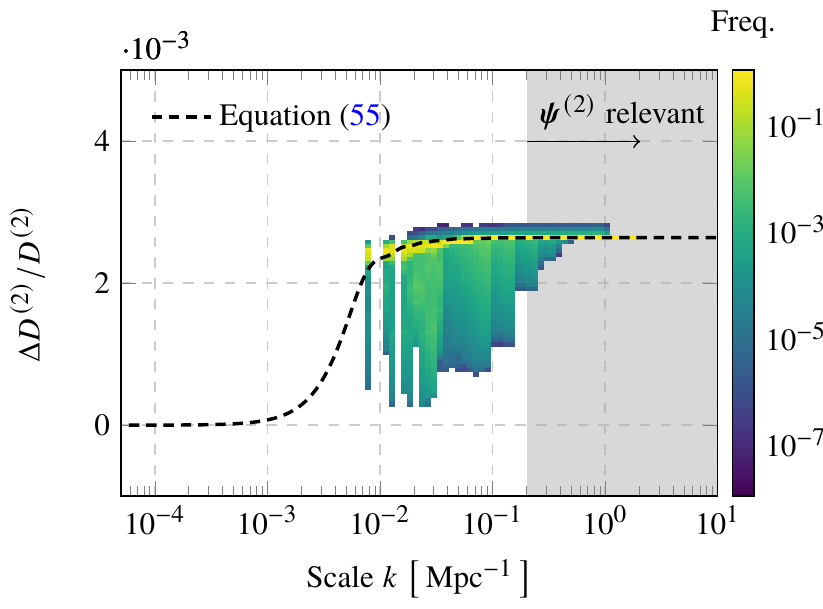}
	\caption{Correction to the $\Lambda$CDM prediction of $D^{(2)}=(3/7)D^2$ for the second-order growth factor, according to the approximate model of Eq.~\eqref{eq:model}, for $\sum m_\nu=0.3$ eV at $z=31$ (dashed line). The colours represent a histogram of the full numerical solution, $D^{(2)}_B(\boldsymbol{k}_1,\boldsymbol{k}_2)$, evaluated on a 6D Fourier space lattice with physical dimension $L=800\text{ Mpc}$ (i.e. $\Delta k=7.85\times 10^{-3}$ $\text{Mpc}^{-1}$), projected onto the $k=\;\rvert\!\rvert\,\boldsymbol{k}_1+\boldsymbol{k}_2\rvert\!\rvert$-axis and normalized per $k$-bin. For the large majority of configurations, the system attains the approximate value. The shaded region indicates the range of scales for which the power spectrum of $\boldsymbol{k}\cdot\bm{\psi}^{(2)}$ is at least $0.01\%$ of that of $\boldsymbol{k}\cdot\bm{\psi}^{(1)}$.}
	\label{fig:EdS_correction}
\end{figure}

\subsection{Limiting solutions}\label{sec:approx}

Having set up the Lagrangian equations for the neutrino-cb fluid model, we are now in a position to look for approximate solutions. The aim is to find expressions for the displacement on large and small scales. In the small-scale limit, neutrinos do not cluster and only contribute to the background expansion as encoded by $g_\infty$. Meanwhile, in the large-scale limit, neutrinos cluster like cold dark matter and one recovers behaviour analogous to $\Lambda$CDM. In both cases, we can find simple solutions in the form of LPT recursion relations \citep{rampf12b,zheligovsky14,matsubara15,rampf15,schmidt21}. These limiting solutions will be used as initial conditions for the numerical integration of the general problem and provide the basis for the recipe of section \ref{sec:ics}.

In this section, we assume that $g_\infty=\text{constant}$, which is exact during matter domination (Eq.~\eqref{eq:early_g_infty}), and a very good approximation in general (Fig. \ref{fig:approx_validity_1}). On large scales, we also have $1+\alpha(k)=1+f_\nu/f_\cb$\footnote{This is not strictly true, since $\delta_\nu > \delta_\cb$ on the largest scales due to the relativistic tail of the neutrino distribution. We ignore this small effect in the current section and in Fig.~\ref{fig:EdS_correction}, but take it into account in section \ref{sec:full}.} and on small scales $1+\alpha(k)=1$. Hence, if all modes involved in the problem are either large or small, we can approximate the convolution with the neutrino response as multiplication by a constant $\beta=1+\alpha(k)$. In such cases, the frame-lagging terms also vanish, as will be confirmed in section \ref{sec:full}. Given these assumptions, \eqref{eq:compact_long} reduces to
\begin{align}
    J^{-1}_{ij}\mathcal{D}_\infty \psi_{i,j} &= \frac{3 \beta g_\infty}{2D_\infty^2}(1-1/J). \label{eq:limit_compact_long}
\end{align}

\noindent
Using the identities $J J^{-1}_{ij} = (1/2)\epsilon_{jkp}\epsilon_{iqr}J_{kq}J_{pr}$ and $J = (1/6)\epsilon_{ijk}\epsilon_{pqr}J_{ip}J_{jq}J_{kr}$, we rewrite \eqref{eq:limit_compact_long} and \eqref{eq:compact_trans} as
\begin{align}
    &\epsilon_{ijk}\epsilon_{pqr}J_{qj}J_{ip}\left[\mathcal{D}_\infty - \frac{\beta g_\infty}{2D_\infty^2}\right]J_{kr} + \frac{3\beta g_\infty}{D_\infty^2} = 0,\\
    &\epsilon_{lpq}J_{qk}\mathcal{D}_\infty\psi_{k,l} = 0.
\end{align}

\noindent
Hence, using $J_{ij} = \delta_{ij} + \psi_{i,j}$ and substituting the expansion \eqref{eq:powers}, we obtain equations for the longitudinal and transverse parts at order $n$ in terms of perturbations of orders $m_1+m_2=n$ (for $n\geq 2$) and $m_1+m_2+m_3=n$ (for $n\geq 3$):
\begin{align}
    &\left[\mathcal{D}_\infty - \frac{3\beta g_\infty}{2D_\infty^2}\right]\nabla\cdot\bm{\psi}^{(n)} =\nonumber\\
    & \;\;\;\; - \sum_{m_1+m_2=n}\epsilon_{ijk}\epsilon_{ipq}\psi_{j,p}^{(m_1)}\left[\mathcal{D}_\infty - \frac{3\beta g_\infty}{4D_\infty^2}\right]\psi_{k,q}^{(m_2)} \label{eq:long} \\
    & \;\;\;\; -\sum_{m_1+m_2+m_3=n}\epsilon_{ijk}\epsilon_{pqr} \frac{1}{2}\psi_{i,p}^{(m_1)}\psi_{j,q}^{(m_2)}\left[\mathcal{D}_\infty - \frac{\beta g_\infty}{2D_\infty^2}\right]\psi_{k,r}^{(m_3)},\nonumber\\
    &\mathcal{D}_\infty\nabla\times\bm{\psi}^{(n)} = \sum_{m_1+m_2=n}\nabla\psi_{i}^{(m_1)}\times\mathcal{D}_\infty\nabla\psi_{i}^{(m_2)}. \label{eq:trans}
\end{align}

\noindent
The first-order equations separate. The longitudinal equation \eqref{eq:long} has the particular time-dependent solution
\begin{align*}
    D^{(1)} = D_\infty^q \;\,\;\,\text{with}\;\,\;\, q = \tfrac{1}{4}\sqrt{4+3g_\infty(8\beta + 3g_\infty -4)}-\tfrac{3}{4}g_\infty + \tfrac{1}{2},
\end{align*}

\noindent
while the transverse equation \eqref{eq:trans} has constant and decaying solutions. Identifying the fastest growing solutions order by order, we find that $\bm{\psi}^{(n)}\propto D_\infty^{nq}$. In particular, we find that the fastest growing solution at second order grows as
\begin{align}
	\frac{D^{(2)}}{D_\infty^{2q}} = \frac{3g_\infty\beta}{4q(2q-1)+3g_\infty(2q-\beta)}. \label{eq:model}
\end{align}

\noindent
Reinserting $\beta=1+\alpha(k)$, we obtain a useful approximation of the magnitude of neutrino effects on the second-order coefficient, relative to the $\Lambda$CDM value of $3/7$. This is shown by the dashed line in Fig.~\ref{fig:EdS_correction} for a model with $\sum m_\nu = 0.3$ eV at $z=31$. We stress that this approximation neglects the non-trivial coupling with the neutrino response in the general case. As we will see in the next section, the second-order solution can be described in full by two kernels, $D^{(2)}_A(\boldsymbol{k}_1,\boldsymbol{k}_2)$ and $D^{(2)}_B(\boldsymbol{k}_1,\boldsymbol{k}_2)$. For most configurations on the 6D Fourier space lattice that we use to generate $N$-body ICs, both $k_1$ and $k_2$ are large and the result is close to the estimate of Eq.~\eqref{eq:model}. However, for cases with one mode large and one mode small or for squeezed configurations with $k=\;\rvert\!\rvert\,\boldsymbol{k}_1+\boldsymbol{k}_2\,\rvert\!\rvert\ll k_1\approx k_2$, the value may depart from this estimate, as shown by the histogram in Fig.~\ref{fig:EdS_correction}. Nevertheless, the figure demonstrates that the large- and small-scale limits provide reasonable bounds on the effect at intermediate scales. Overall, the magnitude of the effect is $\mathcal{O}\big(10^{-3}\big)$, in line with the estimate given in section \ref{sec:disp} for this mass. The figure also demonstrates that the $\Lambda$CDM value of $3/7$ is only reached for $k<10^{-3}\text{ Mpc}^{-1}$, while the second-order potential is important for $k>10^{-1}\text{ Mpc}^{-1}$, reflecting the hierarchy between the neutrino free-streaming scale and the nonlinear scale, $k_\text{fs}\ll k_\text{nl}$, that motivates the approach of section \ref{sec:ics}.

Using $\bm{\psi}^{(n)}\propto D_\infty^{nq}$, we derive recursion relations for the fastest growing solution at order $n\geq2$:
\begin{align}
\begin{split}
    &\nabla\cdot\bm{\psi}^{(n)} = - \sum_{m_1+m_2=n}\frac{1}{2}\left[1-\frac{4m_1m_2q^2}{2nq(nq-1)+3g_\infty(nq-\beta)}\right]\\
    & \;\;\;\;\;\;\;\;\;\;\;\;\;\;\; \times\epsilon_{ijk}\epsilon_{ipq}\psi_{j,p}^{(m_1)}\psi_{k,q}^{(m_2)} \\
    & \;\;\;\;\;\;\;\;\;\;\;\; -\sum_{m_1+m_2+m_3=n}\left[1-\frac{4(m_1m_2+m_2m_3+m_3m_1)q^2}{2nq(nq-1)+3g_\infty(nq-\beta)}\right]\\
    & \;\;\;\;\;\;\;\;\;\;\;\;\;\;\; \times\epsilon_{ijk}\epsilon_{pqr} \frac{1}{6}\psi_{i,p}^{(m_1)}\psi_{j,q}^{(m_2)}\psi_{k,r}^{(m_3)},\\
\end{split} \label{eq:approx_general1}\\
    &\nabla\times\bm{\psi}^{(n)} = \sum_{m_1+m_2=n}\frac{1}{2}\frac{m_2-m_1}{n}\nabla\psi_{i}^{(m_1)}\times\nabla\psi_{i}^{(m_2)}. \label{eq:approx_general2}
\end{align}

\noindent
For the purposes of higher-order ICs, we are primarily interested in deriving corrections to the $\Lambda$CDM coefficients in the small-scale limit with $\beta=q=1$. Reading off coefficients from \eqref{eq:approx_general1}, we find that these can be conveniently expressed in terms of
\begin{align}
    C_n = \frac{(2n+3)g_\infty}{2n+3g_\infty}. \label{eq:asymp_n}
\end{align}

\noindent
Proceeding as in Appendix \ref{sec:3LPT_derivation}, we obtain the 3LPT form given in section \ref{sec:disp}. Combining Eqs.~\eqref{eq:asymp_n} and \eqref{eq:early_g_infty} yields an accurate approximation of $C_n$ in terms of $f_\nu$:
\begin{align}
C_n &= \frac{8(1-f_\nu)(2n+3)}{n(S-1)^2+(S^2-1)} \cong 1 + \frac{2f_\nu n}{5(2n+3)}, \label{eq:cnfact}
\end{align}

\noindent
with $S=\sqrt{1+24(1-f_\nu)}$. For $n=2$, the above expression agrees with that given by \citet{wright17}. The next section is dedicated to relaxing the assumptions on $g_\infty$ and $\alpha(k)$, finding the general solution at second order.

\subsection{General solution}\label{sec:full}

For the general solution, we need to deal with the frame-lagging terms $F(\boldsymbol{q})$. Here, we will follow the approach of \citet{aviles20}. We are interested in solutions at second order. The transverse equation \eqref{eq:compact_trans} only has non-trivial solutions for $n\geq 3$. Therefore, we concentrate on the longitudinal part. We repeat \eqref{eq:compact_long} for convenience:
\begin{align}
	J^{-1}_{ij}\mathcal{D}_\infty \psi_{i,j} &= \frac{3g_\infty}{2D_\infty^2}\left[(1-1/J)*(1+\alpha) + F\right].
\end{align}

\noindent
Using \eqref{eq:jacobian} and $J^{-1}_{ij}=\sum_{n=0}^\infty[(I-J)^n]_{ij}$, we can write this up to second order in the displacement:
\begin{align}
\begin{split}
	&\mathcal{D}_\infty \psi_{i,i} = \psi_{i,j} \mathcal{D}_\infty \psi_{j,i} + \frac{3g_\infty}{2D_\infty^2}\psi_{i,i}*(1+\alpha) \\
	& \;\;\;\;\;\; - \frac{3g_\infty}{2D_\infty^2}\frac{1}{2}\left[\psi_{i,i}\psi_{j,j}+\psi_{i,j}\psi_{j,i}\right]*(1+\alpha) + \frac{3g_\infty}{2D_\infty^2}F^{(2)},
\end{split} \label{eq:full_two}
\end{align}

\noindent
where the second-order frame-lagging terms, $F^{(2)}$, are given in Appendix \ref{sec:FL}. At first order, the displacement admits a growing solution $\bm{\psi}^{(1)}\propto D^{(1)}$ with a growth factor that satisfies
\begin{align}
	\mathcal{D}_\infty D^{(1)} = \frac{3g_\infty}{2D_\infty^2}(1+\alpha)D^{(1)}. \label{eq:lagrange_first_ord}
\end{align}

\noindent
This is simply a reformulation of the Eulerian equation for the first-order growth factor \eqref{eq:growing}. Using the expansion \eqref{eq:powers} in \eqref{eq:full_two} and collecting second-order terms then yields
\begin{align}
\begin{split}
	&\mathcal{D}_\infty\psi^{(2)}_{i,i} = \frac{3g_\infty}{2D_\infty^2}\psi^{(2)}_{i,i}*(1+\alpha) + \psi^{(1)}_{i,j} \mathcal{D}_\infty \psi^{(1)}_{j,i}\\
	&\;\;\;- \frac{3g_\infty}{2D_\infty^2}\frac{1}{2}\left[\psi^{(1)}_{i,i}\psi^{(1)}_{j,j}+\psi^{(1)}_{i,j}\psi^{(1)}_{j,i}\right]*(1+\alpha) + \frac{3g_\infty}{2D_\infty^2}F^{(2)}.
\end{split} \label{eq:second_eq}
\end{align}

\noindent
In Fourier space, each of the quadratic terms in \eqref{eq:second_eq}, including the second-order frame-lagging term, is a convolution of derivatives of $\bm{\psi}^{(1)}(\boldsymbol{k}_1)$ and $\bm{\psi}^{(1)}(\boldsymbol{k}_2)$. Expressing the displacements in terms of potentials as
\begin{align}
    \bm{\psi}^{(1)}=-\nabla\varphi^{(1)}, \;\;\;\;\;\;\;\;\;\;\;\;\;\;    \bm{\psi}^{(2)}=-\nabla\varphi^{(2)},
\end{align}

\noindent
and identifying terms, we thus obtain
\begin{align}
\begin{split}
	\varphi^{(2)} (\boldsymbol{k}) &=  \frac{1}{2}\int_{\boldsymbol{k}_1,\boldsymbol{k}_2}\frac{1}{(ik)^2}\frac{1}{D_1D_2} \varphi^{(1)}(\boldsymbol{k}_1) \varphi^{(1)}(\boldsymbol{k}_2)\\
    &\;\;\;\;\;\;\;\; \times \bigg[D^{(2)}_A(\boldsymbol{k}_1,\boldsymbol{k}_2)k_1^2k_2^2 -  D^{(2)}_B(\boldsymbol{k}_1,\boldsymbol{k}_2)k_{12}^2\bigg],
\end{split} \label{eq:second_pot}
\end{align}

\noindent
where $\int_{\boldsymbol{k}_1,\boldsymbol{k}_2} = \int\mathrm{d}\boldsymbol{k}_1\mathrm{d}\boldsymbol{k}_2(2\pi)^{-6}\delta^{(3)}(\boldsymbol{k}_1+\boldsymbol{k}_2-\boldsymbol{k})$ and $k_{12}=\boldsymbol{k}_1\cdot\boldsymbol{k}_2$ and $D_i=D^{(1)}(k_i)$ for $i=1,2$. Notice the similarity of this equation with Eq.~\eqref{eq:approx_2lpt}. The difference is that the two terms now have distinct scale- and time-dependent coefficients satisfying
\begin{align}
	\mathcal{D}_\infty D^{(2)}_A &= \frac{3g_\infty}{2D_\infty^2} \left(1+\alpha(k)\right) D^{(2)}_A + \frac{3g_\infty}{2D_\infty^2} (1+A) D_1D_2, \label{eq:diffeq_A}\\
	\mathcal{D}_\infty D^{(2)}_B &= \frac{3g_\infty}{2D_\infty^2} \left(1+\alpha(k)\right) D^{(2)}_B + \frac{3g_\infty}{2D_\infty^2}
    (1+B) D_1D_2, \label{eq:diffeq_B}
\end{align}

\noindent
where the functions $A$ and $B$ are given by
\begin{align}
	A(k,k_1,k_2) &= \alpha(k) + \left[\frac{\alpha(k)-\alpha(k_2)}{k_1^2} + \frac{\alpha(k)-\alpha(k_1)}{k_2^2}\right]k_{12},\\
	B(k,k_1,k_2) &= \alpha(k_1) + \alpha(k_2) - \alpha(k),
\end{align}

\noindent
for $k=\;\rvert\!\rvert\,\boldsymbol{k}_1+\boldsymbol{k}_2\rvert\!\rvert$. The terms in square brackets correspond to the frame-lagging terms. In the small-scale limit with $k,k_1,k_2\gg k_\text{fs}$, we have $A=B=0$. Hence, $D^{(2)}_A=D^{(2)}_B$ and \eqref{eq:second_pot} factorizes as in Eq.~\eqref{eq:approx_2lpt}. Similarly, in the large-scale limit with $k,k_1,k_2\ll k_\text{fs}$, we obtain again the approximate form described in section \ref{sec:approx} with $A=B\approx f_\nu/f_\cb$. In both limits, the frame-lagging terms drop out, as anticipated. Intermediate configurations will deviate from the asymptotic solutions, as was already discussed in section \ref{sec:approx} and shown in Fig.~\ref{fig:EdS_correction}.

For the numerical solution, we begin the integration at a time when the non-relativistic neutrino fraction is 50\%. For the fiducial neutrino mass, $\sum m_\nu=0.3$ eV, this corresponds to $z=187$. We integrate Eqs.~\eqref{eq:lagrange_first_ord} for the first-order growth factor and (\ref{eq:diffeq_A}-\ref{eq:diffeq_B}) for the second-order kernels, using the approximate model of Eq.~\eqref{eq:model} as initial conditions. The results, projected onto the $k$-axis, are shown in Fig.~\ref{fig:EdS_correction}. When generating 2LPT particle initial conditions, we begin by generating a realisation of the back-scaled first-order potential, $\varphi^{(1)}$. We then perform the convolution integral of Eq.~\eqref{eq:second_pot} explicitly, interpolating from tables of $D^{(2)}_{A,B}(k,k_1,k_2)$. To ensure completion in a reasonable time frame, we impose cut-offs at $k_1\leq k_\text{cut}$ and $k_2\le k_\text{cut}$. We performed convergence tests to ensure that the results are independent of the cut-off scale, finding that a cut-off at $k_\text{cut}=1\text{ Mpc}^{-1}$ was more than adequate for the resolutions considered in this paper.

\begin{figure}
	\normalsize
	\centering
    \includegraphics{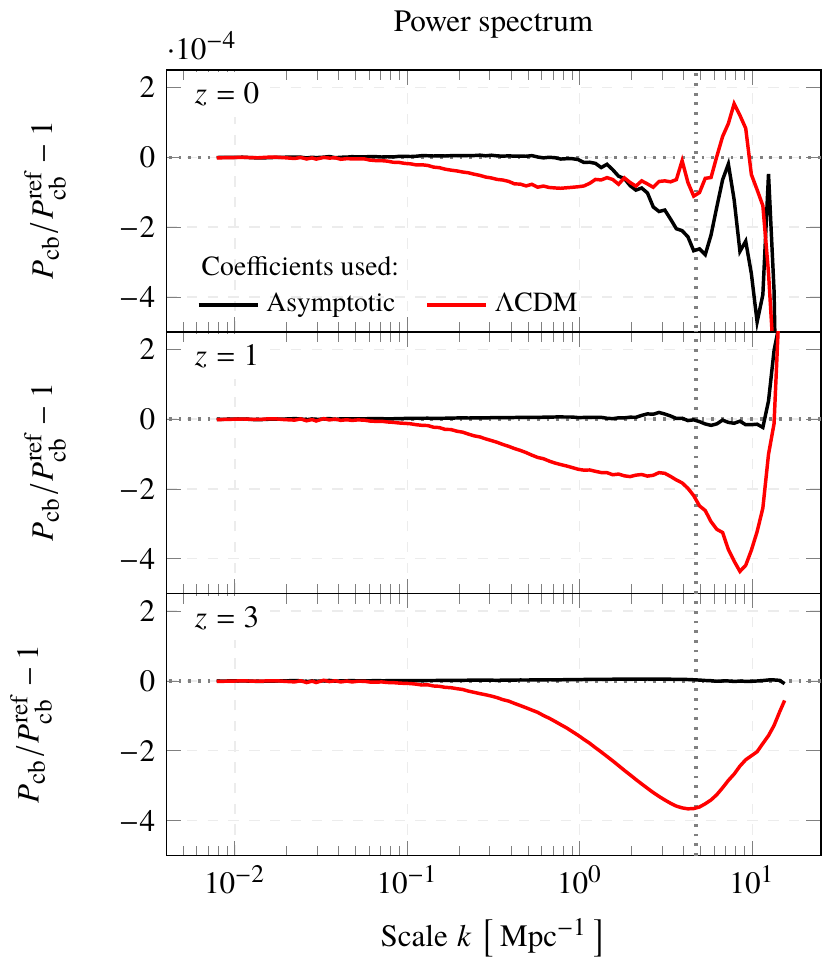}
    \caption{Impact of approximation schemes for the second-order potential on the CDM \& baryon power spectrum. The reference run used initial conditions based on a numerical calculation of the scale-dependent 2LPT kernels, $D^{(2)}_A(\boldsymbol{k}_1,\boldsymbol{k}_2)$ and $D^{(2)}_B(\boldsymbol{k}_1,\boldsymbol{k}_2)$. In the asymptotic approximation (black), we use Eqs.~\eqref{eq:3LPT} and \eqref{eq:veloc}, but truncate third-order terms. In the $\Lambda$CDM approximation (red), we additionally set $C_2=1$. The vertical dotted line is the particle Nyquist frequency}
    \label{fig:approx_impact}
\end{figure}

\begin{table}
   \centering
   \caption[caption]{Description of the gravitational parameters used by \textsc{swift} for the $N_\cb=600^3$ (low-res) and $N_\cb=1200^3$ (high-res) simulations.}
   \begin{tabular}{  l  l l }
       \hline
       \hline
       \textbf{Parameter} & \textbf{Low-res} $\;\;\;\;\;\;$ & \textbf{High-res} \\
       \hline
        \texttt{mesh\_side\_length} & 512 & 1024\\
        \texttt{MAC} & adaptive & adaptive\\
        \texttt{epsilon\_fmm} & 0.001 & 0.001\\
        \texttt{eta} & 0.025 & 0.025\\
        \texttt{theta\_cr} & 0.7 & 0.7 \\
        \texttt{use\_tree\_below\_softening} & 1 & 1\\
        \texttt{comoving\_DM\_softening} & 0.0533333 & 0.0266667\\
        \texttt{max\_physical\_DM\_softening} & 0.0533333 & 0.0266667\\
        \texttt{comoving\_nu\_softening} & 0.0533333 & 0.0266667\\
        \texttt{max\_physical\_nu\_softening} & 0.0533333 & 0.0266667\\
        \hline
       \hline
   \end{tabular}
   \label{tab:sim_params}
\end{table}

\begin{table}
	\centering
	\caption[caption]{Description of the simulations. The listed particle mass, $m_p$, refers to the cb particles. The neutrino fraction is listed as $f_\nu=\Omega_\nu/(\Omega_\text{cb} + \Omega_\nu)$. All simulations used the same random phases in an $L=800\text{ Mpc}$ cube.}
	\begin{tabular}{  l  l  c  l  l  l l }
		\hline
		\hline
		ICs & $z_i$ & $N_\cb$ & $m_p\;\big[\,M_\odot\,\big]$ & $N_\nu$ & $\sum m_\nu$ & $f_\nu$\\
		\hline
        ZA & $127$ & $1200^3$ & $1.14\times10^{10}$ & $600^3$ & $0.30$ eV & $0.023$\\
        ZA & $63$ & $1200^3$ & $1.14\times10^{10}$ & $600^3$ & $0.30$ eV & $0.023$\\
		ZA & $31$ & $1200^3$ & $1.14\times10^{10}$ & $600^3$ & $0.30$ eV & $0.023$\\
        2LPT & $31$ & $1200^3$ & $1.14\times10^{10}$ & $600^3$ & $0.30$ eV & $0.023$\\
        3LPT & $31$ & $1200^3$ & $1.14\times10^{10}$ & $600^3$ & $0.30$ eV & $0.023$\\
        \hline
        2LPT & $31$ & $1200^3$ & $1.17\times10^{10}$ & $600^3$ & $0.00$ eV & $0.0$\\
        2LPT & $31$ & $1200^3$ & $1.14\times10^{10}$ & $600^3$ & $0.30$ eV & $0.023$\\
        2LPT & $127$ & $1200^3$ & $1.17\times10^{10}$ & $600^3$ & $0.00$ eV & $0.0$\\
        2LPT & $127$ & $1200^3$ & $1.14\times10^{10}$ & $600^3$ & $0.30$ eV & $0.023$\\
        \hline
        ZA & $31$ & $600^3$ & $9.34\times10^{10}$ & $600^3$ & $0.00$ eV & $0.0$\\
        ZA & $31$ & $600^3$ & $9.23\times10^{10}$ & $600^3$ & $0.15$ eV & $0.011$\\
        ZA & $31$ & $600^3$ & $9.12\times10^{10}$ & $600^3$ & $0.30$ eV & $0.023$\\
        2LPT & $31$ & $600^3$ & $9.24\times10^{10}$ & $600^3$ & $0.00$ eV & $0.0$\\
        2LPT & $31$ & $600^3$ & $9.23\times10^{10}$ & $600^3$ & $0.15$ eV & $0.011$\\
        2LPT & $31$ & $600^3$ & $9.12\times10^{10}$ & $600^3$ & $0.15$ eV & $0.011$\\
        3LPT & $31$ & $600^3$ & $9.34\times10^{10}$ & $600^3$& $0.00$ eV & $0.0$\\
        3LPT & $31$ & $600^3$ & $9.23\times10^{10}$ & $600^3$ & $0.15$ eV & $0.011$\\
        3LPT & $31$ & $600^3$ & $9.12\times10^{10}$ & $600^3$ & $0.30$ eV & $0.023$\\
		\hline
        2LPT & $63$ & $600^3$ & $9.24\times10^{10}$ & $600^3$ & $0.00$ eV & $0.0$\\
        2LPT & $63$ & $600^3$ & $9.23\times10^{10}$ & $600^3$ & $0.15$ eV & $0.011$\\
        2LPT & $63$ & $600^3$ & $9.12\times10^{10}$ & $600^3$ & $0.15$ eV & $0.011$\\
        2LPT & $127$ & $600^3$ & $9.24\times10^{10}$ & $600^3$ & $0.00$ eV & $0.0$\\
        2LPT & $127$ & $600^3$ & $9.23\times10^{10}$ & $600^3$ & $0.15$ eV & $0.011$\\
        2LPT & $127$ & $600^3$ & $9.12\times10^{10}$ & $600^3$ & $0.15$ eV & $0.011$\\
        \hline
		\hline
	\end{tabular}
	\label{tab:sims}
\end{table}

\section{Results}\label{sec:results}

We will now discuss the power spectra, bispectra, and halo mass functions of massive neutrino simulations with different ICs. We introduce our simulation suite in section \ref{sec:sims}. We then consider the impact of different approximation schemes for the second-order kernels in section \ref{sec:validate} and follow it up with a comparison of Zel'dovich (ZA), 2LPT, and 3LPT ICs at various starting redshifts in section \ref{sec:lpt_order}. Finally, we consider the impact of ICs on the suppression of the power spectrum as a function of neutrino mass in section \ref{sec:mass_choices}.

\subsection{Simulations}\label{sec:sims}

We use the cosmological hydrodynamics code \textsc{swift} \citep{schaller16,schaller18}, which uses task-based parallelism, asynchronous communication, fast neighbour finding, and vectorised operations to achieve significant speed-ups. The code uses the Fast Multipole Method (FMM) for short-range gravitational forces and the Particle Mesh method for long-range forces. Neutrinos are modelled as a separate particle species. We employ the $\delta f$ method to suppress the effects of shot noise \citep{elbers20} and generate neutrino particle initial conditions by integrating geodesics from high redshift using our \textsc{FastDF} code. Additionally, we use fixed initial conditions to limit cosmic variance \citep{angulo16}. Apart from the neutrino mass, we use cosmological parameters based primarily on Year 3 results from the Dark Energy Survey \citep{porredon21} and Planck 2018 \citep{planck18}. Our choice of parameters is $(h, \Omega_\text{m}, \Omega_\text{b}, A_s, n_s)=(0.681, 0.306, 0.0486, 2.09937\times10^{-9}, 0.967)$, with different choices for the neutrino density $\Omega_\nu$. The parameters used by the gravity solver are listed in table \ref{tab:sim_params} and an overview of the simulations is given in table \ref{tab:sims}.

There is a subtle point regarding comparisons between simulations with and without massive neutrinos. Codes like \textsc{swift} employ a multipole acceptance criterion to determine when the multipole approximation is sufficiently accurate to be used without further refinement. The adaptive criterion used for the runs in this paper is based on error analysis of forces on test particles. This means that the accuracy of the $N$-body calculation depends on the number of particles contained in any given volume. When comparing two runs with equal numbers of dark matter particles, one with neutrinos and the other without, all other things being equal, forces will be calculated more accurately in the run with neutrinos. To account for this difference, we included an equal number of massless `spectator' neutrino particles in the $f_\nu = 0$ runs, with velocities corresponding to $m_\nu=0.05$ eV neutrinos. These particles contribute no forces and only affect the $N$-body simulation through the multipole acceptance criterion, ensuring that the accuracy of the massless runs is comparable to that of the massive neutrino runs. Such massless runs are considered in section \ref{sec:mass_choices}.

\subsection{Validation of approximate treatment}\label{sec:validate}

To validate our approach, we compare three different implementations of 2LPT, based on the following models:
\begin{enumerate}[widest=99,itemindent=*,leftmargin=0pt]
    \item The asymptotic model of section \ref{sec:ics}
    \item A model with $\Lambda$CDM coefficients
    \item A reference model with scale-dependent effects
\end{enumerate}

\noindent
The first order displacements and velocities are identical in each of the approaches, obtained from the back-scaled linear power spectrum at $z=0$. In the asymptotic scheme, we use Eqs.~\eqref{eq:3LPT} and \eqref{eq:veloc}, but truncate the 3LPT terms. In the $\Lambda$CDM approximation, we additionally set $C_2=1$, which corresponds to neglecting neutrino effects at second order. Finally, we compare these two approximate methods with a reference run that relied on a numerical calculation of the scale-dependent 2LPT kernels, $D^{(2)}_A(\boldsymbol{k}_1,\boldsymbol{k}_2)$ and $D^{(2)}_B(\boldsymbol{k}_1,\boldsymbol{k}_2)$. With respect to Fig.~\ref{fig:EdS_correction}, the asymptotic approximation corresponds to using the small-scale limit, the $\Lambda$CDM approximation corresponds to the large-scale limit, and the reference run corresponds to the underlying histogram. We use simulations with side length $L=800\text{ Mpc}$ and $N_\cb=1200^3$ particles.

Fig.~\ref{fig:approx_impact} shows the impact of these approximations on the power spectrum of the evolved CDM \& baryon density field. The differences are most evident at $z=3$ (bottom panel). On the largest scales, $k<0.05\text{ Mpc}^{-1}$, nonlinear corrections are small and all simulations agree to machine precision. For $k>0.05\text{ Mpc}^{-1}$, the $\Lambda$CDM simulation systematically underestimates clustering with a maximum error of $0.04\%$ at $k=4\text{ Mpc}^{-1}$. For the asymptotic run, the error is two orders of magnitude smaller over the same scales. Between $z=31$ and $z=3$, the evolution is virtually identical in the asymptotic and reference runs, but we begin to see some noise in the ratio on the smallest scales at $z=1$ (middle panel). These perturbations continue to grow until $z=0$ (top panel), where we find a scatter of $2\times10^{-4}$ for $k>1\text{ Mpc}^{-1}$ in both the asymptotic/reference and $\Lambda$CDM/reference ratios. It is hard to attribute this noise to any particular run as the power spectrum on these scales is increasingly determined by the internal structure of poorly resolved halos. On larger scales, $k<1\text{ Mpc}^{-1}$, the asymptotic run performs extremely well with errors below $10^{-5}$, while the systematic deficit in the $\Lambda$CDM run persists.

These results demonstrate that, at second order, the effect of the suppressed neutrino perturbations can be absorbed into a scale-independent factor $C_2$ and that further scale-dependent neutrino effects are negligible as far as initial conditions are concerned. We expect that this continues to hold for third-order corrections, which are confined to even smaller scales. Including the correction factor $C_2$ is clearly superior to simply using the $\Lambda$CDM coefficient. However, we also observe that this higher-order neutrino effect is below $0.1\%$, and therefore beyond the sensitivity of current experiments. Hence, we conclude that for most purposes both the $\Lambda$CDM approximation and the asymptotic approximation are justified.

 \begin{figure}
 	\normalsize
 	\centering
    \includegraphics{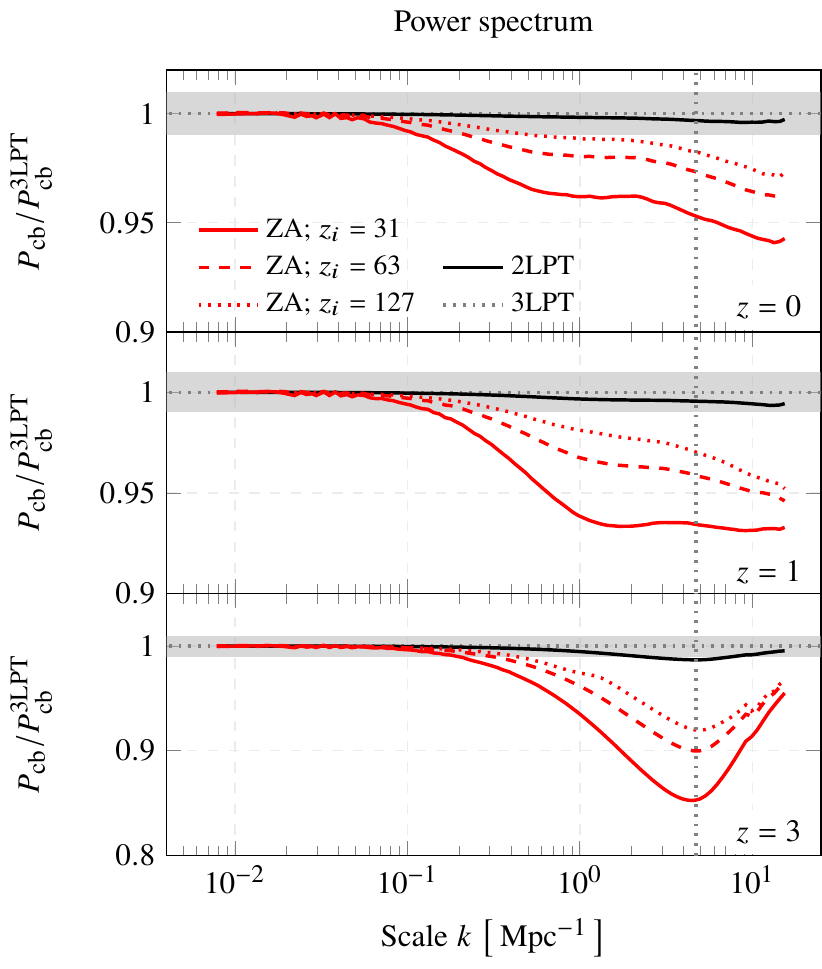}
    \caption{Impact of starting time and LPT order on the CDM \& baryon power spectrum. The reference simulation used 3LPT and both it and the 2LPT simulation were started at $z_i=31$. The shaded area is $1\%$ and the vertical dotted line is the particle Nyquist frequency.}
    \label{fig:order_choice}
 \end{figure}

\subsection{Choice of LPT order and starting time}\label{sec:lpt_order}

\begin{figure}
	\normalsize
	\centering
    \includegraphics{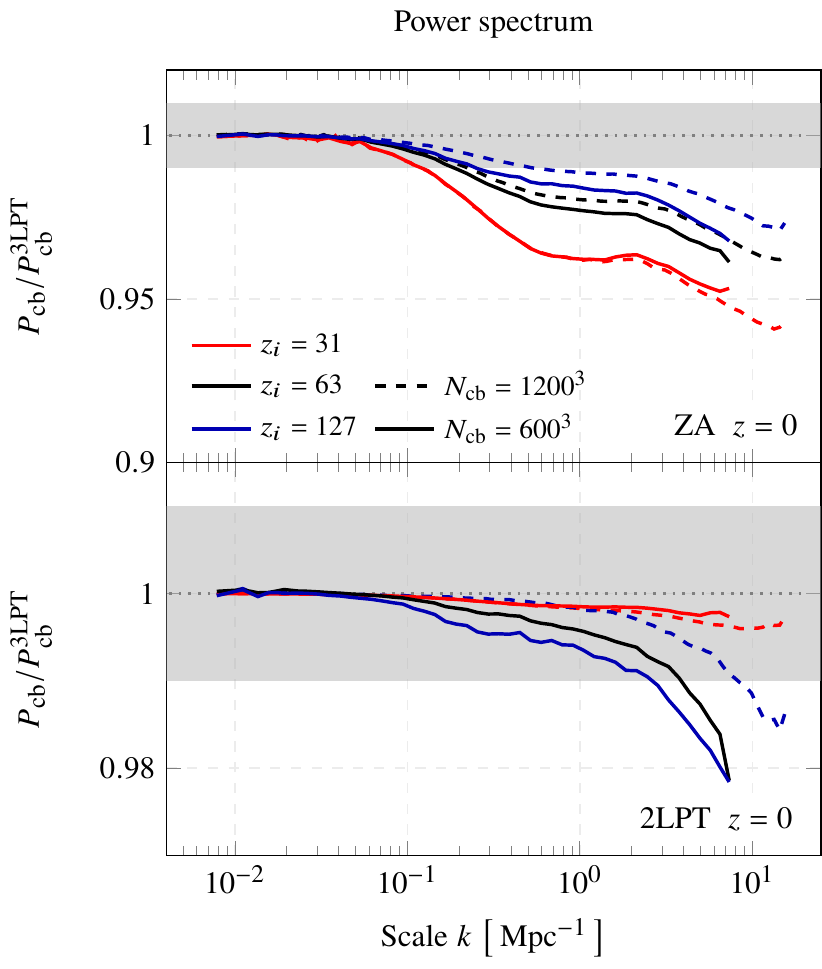}
    \caption{Impact of starting time and resolution on the CDM \& baryon power spectrum. The simulations are compared against 3LPT runs with the same resolution ($N_\cb=600^3$ or $N_\cb=1200^3$), started at $z_i=31$. The shaded area is $1\%$. Not all combinations were tested.}
    \label{fig:starting_time}
\end{figure}

We are now in a position to study the effects of LPT order and starting time on massive neutrino simulations, using the asymptotic approximation. Fig.~\ref{fig:order_choice} shows the late-time power spectrum for simulations with $L=800\text{ Mpc}$ and $N_\cb=1200^3$ particles, comparing in the first instance Zel'dovich (solid red) and 2LPT (solid black) with 3LPT (dotted gray) as a baseline. All three runs were started at $z_i=31$. The most striking observation is that the differences are much larger than those shown in Fig.~\ref{fig:approx_impact}. This means that using higher-order LPT in some fashion is more important than getting the details right. Next, we find per cent agreement between 2LPT and 3LPT over the entire range of scales probed for $z\leq 1$ and approximately a $1\%$ error at $z=3$ for $k>2\text{ Mpc}^{-1}$. We also find that the Zel'dovich approximation performs very poorly with errors of $(4,\, 7,\, 15)\%$ for $k>1\text{ Mpc}^{-1}$ at $z=(0,\,1,\,3)$. This well-known fact \citep{crocce06} has motivated practitioners to start Zel'dovich simulations at higher redshifts, when truncation errors are smaller. We demonstrate this with Zel'dovich runs started at $z_i=63$ (dashed, red) and $z_i=127$ (dotted, red). While the agreement with the higher-order runs improves, we still find per cent agreement only up to $k=0.4\text{ Mpc}^{-1}$. Moreover, starting earlier introduces inaccuracies of a different sort. To see this, we repeat the exercise at a lower resolution with $N_\cb=600^3$ particles. The resulting power spectra at $z=0$ are shown in Fig.~\ref{fig:starting_time}, with Zel'dovich runs compared against 3LPT in the top panel. We observe that for runs started at $z_i=31$ (red), the error is almost independent of resolution. However, for earlier starts at $z=63$ (black) and $z=127$ (blue), the lower resolution runs increasingly underestimate the power spectrum on small scales. This shows that while truncation errors decrease, resolution effects increase as simulations are started earlier. The pattern reverses for 2LPT (bottom panel), with earlier starts performing worse than later starts. This can easily be explained by the fact that truncation errors are much smaller for 2LPT, such that the effect of increasing discreteness errors dominates. We confirm the finding of \citet{michaux21} that the size of discreteness errors is independent of LPT order. This demonstrates that, at fixed resolution and LPT order, starting earlier does not guarantee convergence onto the higher-order solution. As was the case for truncation errors, discreteness errors are much larger at $z=1,\,3$ (not shown).

\begin{figure}
	\normalsize
	\centering
    \includegraphics{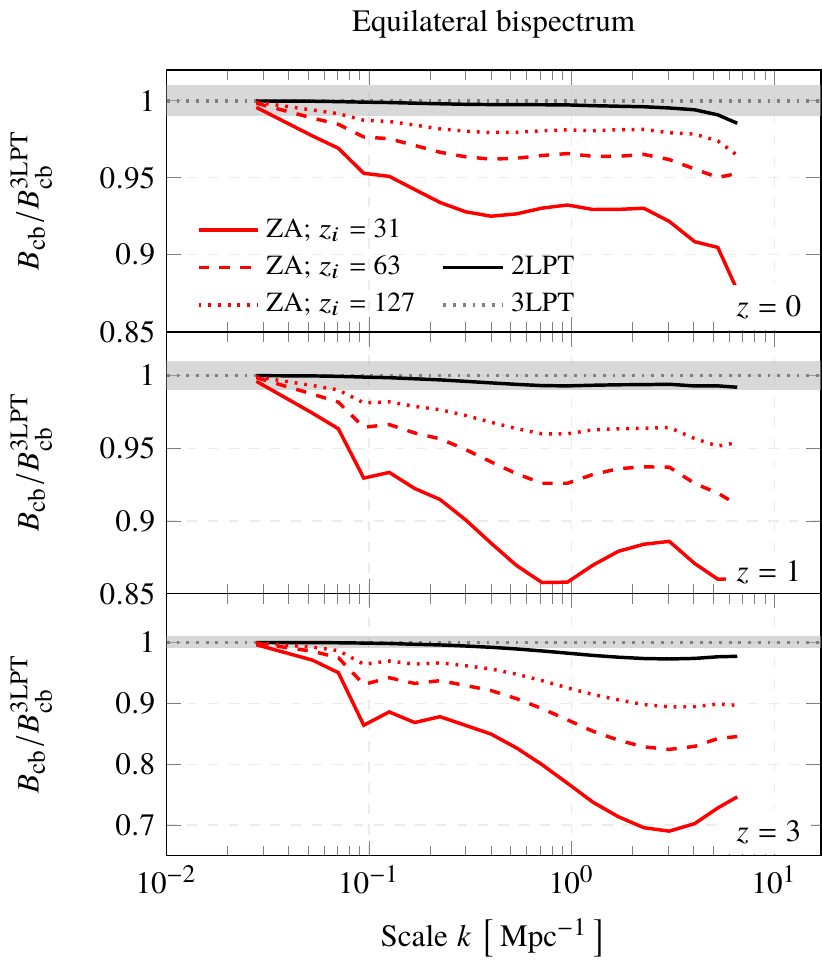}
    \caption{Impact of starting time and LPT order on the equilateral bispectrum of CDM \& baryon density perturbations at late times. The reference simulation used 3LPT and both it and the 2LPT simulation were started at $z_i=31$. All runs used $N_\cb=1200^3$ particles. The shaded area is $1\%$.}
    \label{fig:order_choice_bispec}
\end{figure}

We also consider three-point statistics, which are sensitive to transients from initial conditions \citep{crocce06} and an interesting probe of neutrino physics \citep{chiang18,ruggeri18,hahn20}. For the equilateral bispectrum, $B(k)=B(k_1,k_2,k_3)$ with $k=k_1=k_2=k_3$, shown in Fig.~\ref{fig:order_choice_bispec} at late times, the same pattern is broadly repeated as for the power spectrum. However, errors are approximately twice as large as for the power spectrum. In detail, we again find per cent agreement between 2LPT and 3LPT for $z\leq 1$ with larger errors on small scales at $z=3$. For the Zel'dovich runs, we find significant errors compared to 3LPT, even when starting at $z=127$, with per cent agreement only up to $k=0.1\text{ Mpc}^{-1}$ at $z=0$, and not even there for $z\geq1$.

\begin{figure}
	\normalsize
	\centering
    \includegraphics{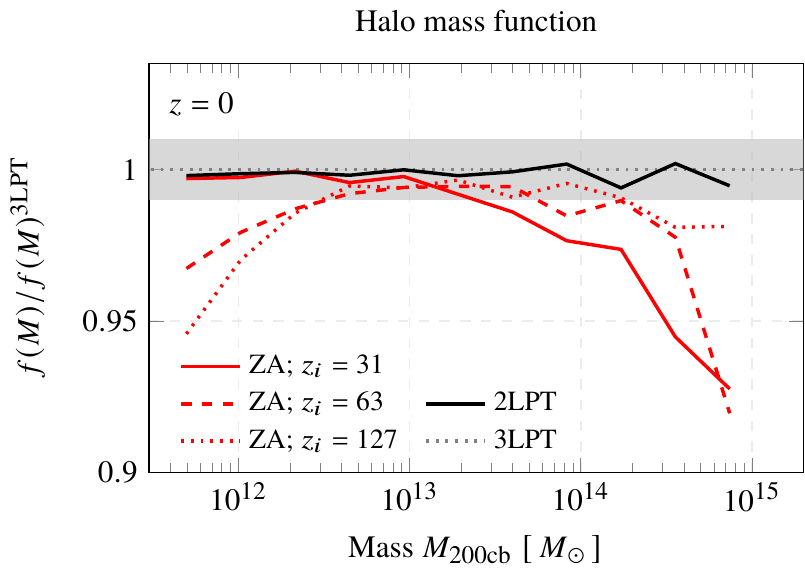}
    \caption{Impact of starting time and LPT order on the halo mass function, $f(M)=\mathrm{d}n/\mathrm{d}\log M$, at $z=0$ for the $N_\cb=1200^3$ runs. The reference simulation used 3LPT and was started at $z_i=31$. The shaded area is $1\%$.}
    \label{fig:hmf}
\end{figure}

Finally, we compare the halo mass function at $z=0$. Halos are identified with \textsc{VELOCIraptor} \citep{elahi19} using a 6D friends-of-friends algorithm applied to the cb particles. Spherical overdensity masses are computed within spheres for which the density equals 200 times the mean CDM \& baryon density $\bar{\rho}_\cb$. The reason for using $\bar{\rho}_\cb$ instead of the total mass density $\bar{\rho}_\text{m}$ is that it is this cold density field that produces universal and unbiased results in halo model calculations \citep{ichiki12,castorina14,massara14}. The results are shown Fig.~\ref{fig:hmf}. We once again find per cent agreement between 2LPT and 3LPT over the entire mass range, but large errors for the Zel'dovich runs. There is an interesting pattern in the Zel'dovich error as the starting time is varied. For late starts (solid red), the simulation agrees well at the low-mass end but underestimates the number of very massive, $10^{15}M_\odot$, halos by more than 7\%. This can be understood in terms of the deficit of power seen also in Fig.~\ref{fig:order_choice}, resulting in a suppressed growth of large structures. Meanwhile, for early starts (dotted and dashed red), the agreement at the high-mass end improves like the small-scale power spectrum. However, the number of low-mass halos decreases by a similar factor, likely due to discreteness errors. This seems to be broadly consistent with the $\Lambda$CDM results of \citet{michaux21}, but not with \citet{nishimichi19} who find little dependence on starting time at $z=0$.

\subsection{Dependence on neutrino mass}\label{sec:mass_choices}

Thus far, we have focused on a single neutrino mass of $\sum m_\nu = 0.3$ eV. However, it is of great interest to determine the effect of initial conditions on the suppression of the power spectrum for different neutrino masses. We consider three cases:
\begin{enumerate}[widest=99,itemindent=*,leftmargin=0pt]
    \item massless neutrinos
    \item degenerate $\sum m_\nu = 0.15$ eV neutrinos ($f_\nu=0.011$),
    \item degenerate $\sum m_\nu = 0.30$ eV neutrinos ($f_\nu=0.023$).
\end{enumerate}

\noindent
In each case, we adjust $\Omega_\text{cdm}$ to keep the total matter density $\Omega_\text{m}$ fixed. We primarily use lower resolution simulations with $N_\cb=600^3$ particles in an $L=800\text{ Mpc}$ cube.

\begin{figure*}
	\normalsize
	\centering
    \subfloat{
    	\includegraphics{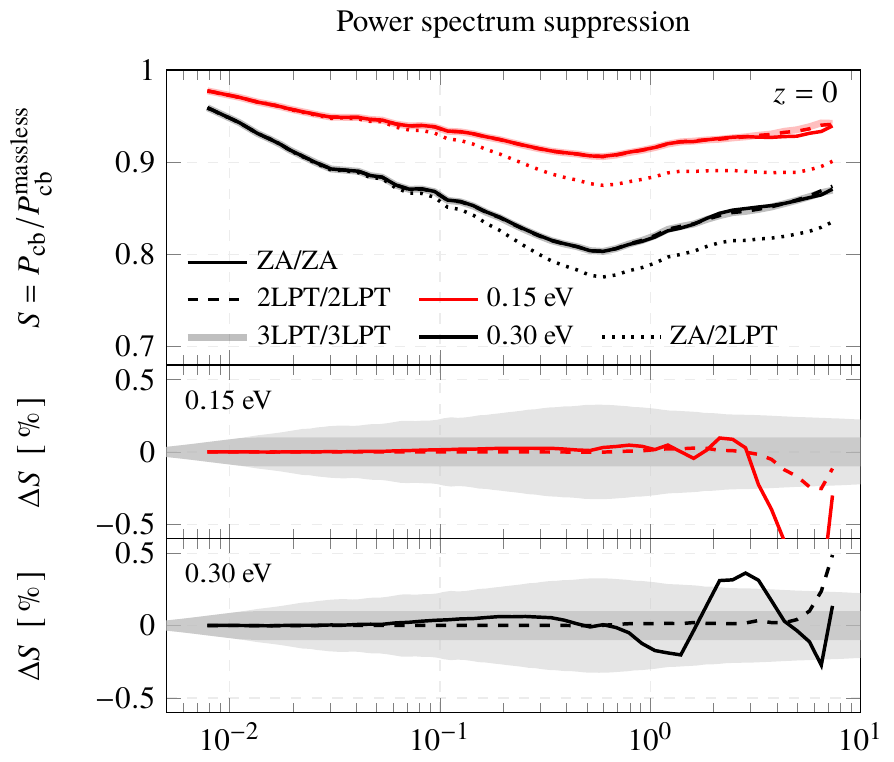}
    }\hspace{-2em}
    \subfloat{
    	\includegraphics{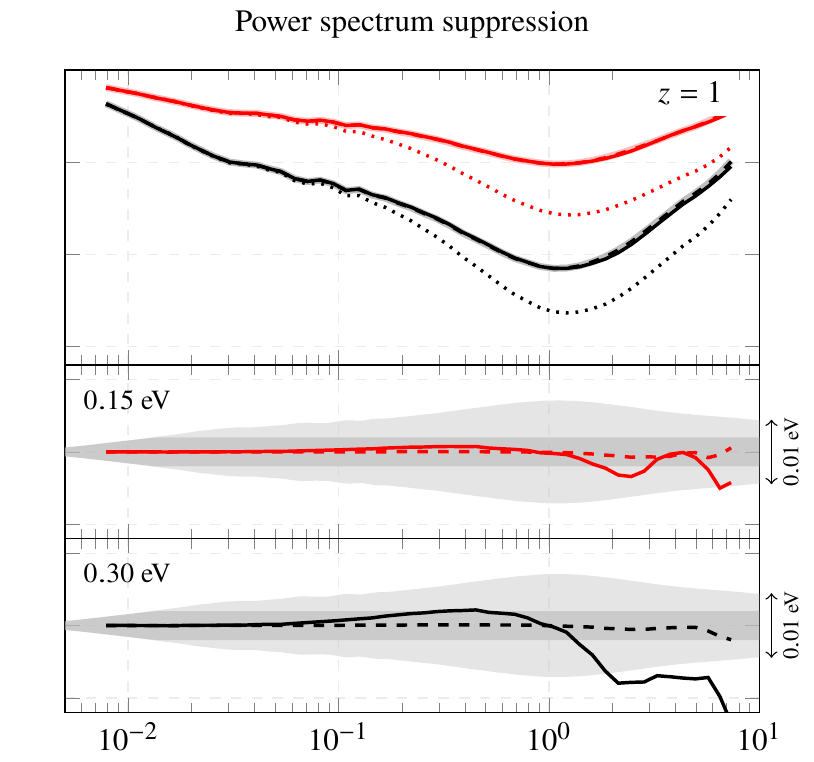}
    }
    \caption{Impact of LPT order on the suppression of the CDM \& baryon power spectrum for different choices of A/B, where A is the LPT order of the massive neutrino run and B the LPT order of the massless run. The neutrino masses are $\sum m_\nu = 0.15$ eV (red) and $\sum m_\nu = 0.30$ eV (black). The bottom panels show the suppression relative to 3LPT/3LPT, with shaded areas representing a $\pm0.005$ eV shift (light) or a constant $0.1\%$ error (dark) where this is smaller.}
    \label{fig:mass_dependence_lpt_order}
\end{figure*}

First, we consider the effect of LPT order. In Fig.~\ref{fig:mass_dependence_lpt_order}, we show the suppression of the CDM \& baryon power spectrum relative to the massless case, comparing ZA/ZA (solid), 2LPT/2LPT (dashed), and 3LPT/3LPT (shaded). Evidently, it is crucial to compare like with like simulation, keeping the LPT order and starting redshift the same. Not doing so introduces large errors in the ratio, as might be expected from the fixed neutrino mass results discussed above. We illustrate this by including a dotted line for the ZA/2LPT ratio, which is clearly off the mark. However, even when comparing like with like, we find a residual error that is proportional to the neutrino mass, rises with $k$, and peaks around the turn-over of the spoon. This feature is most clearly visible at $z=1$ for ZA, with a maximum error of $0.05f_\nu$. The effect is already present in the initial conditions and can be explained by a mass-dependent suppression of nonlinear terms. As virialized structures grow, both the turn-over of the spoon and the peak of the error move to larger scales. At $z=0$, the error is $0.025f_\nu$ around $k=0.3\text{ Mpc}^{-1}$. On smaller scales, we see a scatter of order $0.5\%$, treading outside the scale-dependent error bars that correspond to a $\pm0.005$ eV shift in $\sum m_\nu$. For 2LPT, both the systematic effect and the noise are greatly suppressed, resulting in $0.1\%$-level agreement with 3LPT even at early times.

\begin{figure*}
	\normalsize
	\centering
    \subfloat{
    	\includegraphics{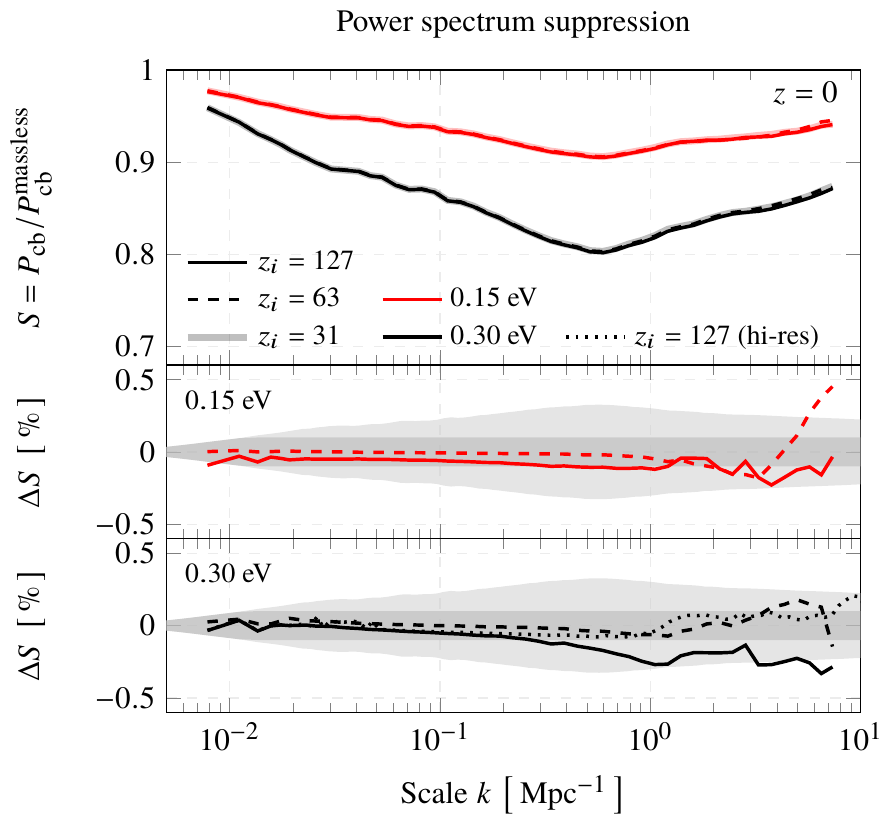}
    }\hspace{-2em}
    \subfloat{
    	\includegraphics{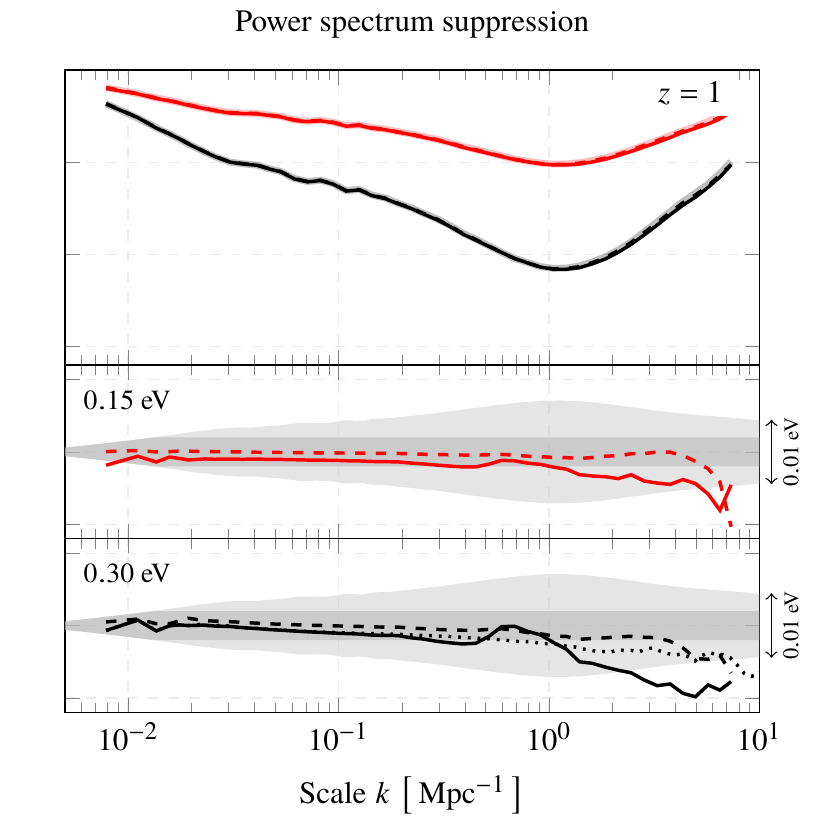}
    }
    \caption{Impact of starting redshift and resolution on the suppression of the CDM \& baryon power spectrum. The neutrino masses are $\sum m_\nu = 0.15$ eV (red) and $\sum m_\nu = 0.30$ eV (black). The bottom panels show the suppression relative to runs with the same resolution, but started at $z_i=31$. The shaded areas represent a $\pm0.005$ eV shift (light) and a constant $0.1\%$ error (dark) where this is smaller. All simulations used 2LPT initial conditions.}
    \label{fig:mass_dependence_time}
\end{figure*}

Next, we consider the effect of the starting time of the simulation. In Fig.~\ref{fig:mass_dependence_time}, we show the suppression for simulations with 2LPT ICs started at $z=127$ (solid), $z=63$ (dashed), $z=31$ (shaded). Once again, we compare like with like simulations. Even so, we find a small residual effect with earlier starts overestimating the suppression. The differences between $z=31$ and $z=63$ are minimal for both neutrino masses. However, starting at $z=127$ results in $(0.1,\,0.2)f_\nu$ errors at $z=(0,\,1)$ for $k>1\text{ Mpc}^{-1}$. These errors once again exceed the threshold for a $\pm0.005$ eV shift in $\sum m_\nu$. Based on the discussion above, and given that we are using 2LPT, we expect that truncation errors are small at both redshifts. This suggests that the differences are caused by resolution effects, which grow in importance with the starting redshift. To test this, we repeated some of the simulations at a higher resolution with $N_\cb = 1200^3$ particles, starting at $z=127$ and $z=31$. The ratio is shown as a dotted line in the bottom panels of Fig.~\ref{fig:mass_dependence_time}. The agreement between the early and late starts improves to $0.1\%$ up to $k=10\text{ Mpc}^{-1}$ at $z=0$, comparable to the low-resolution $z=63$ start. However, the suppression is still slightly overestimated at $z=1$.

One possible alternative explanation is that errors could be introduced by the back-scaling procedure (section \ref{sec:TF}). To test this hypothesis, we repeated some of the simulations with ``forward'' ICs, as in \citet{elbers20}. We found nearly identical results for these runs, ruling out this explanation. Another possibility is that the errors could be the result of shot noise, since we use a particle-based implementation of neutrino perturbations. However, this is unlikely as the differences already appear at high redshift when shot noise is highly suppressed due to our use of the $\delta f$ method. Finally, one might expect differences due to relativistic effects that are increasingly important for earlier starts. Once again, this is unlikely since relativistic effects would appear on the largest scales, where the differences shown in Fig.~\ref{fig:mass_dependence_time} are minimal. Since the error decreases for the higher resolution runs, discreteness effects likely account for the majority of the difference, with massive neutrino simulations being more sensitive to such errors, due to the suppressed growth of structure. Late starts can be utilized to minimize the effect of particle resolution, as shown in Fig.~\ref{fig:starting_time}.

\section{Discussion}\label{sec:conclusion}

We have investigated the use of higher-order Lagrangian initial conditions (ICs) for cosmological simulations with massive neutrinos. We solved the fluid equations for a neutrino-CDM-baryon model with approximate time-dependence in the large- and small-scale limits, finding that higher-order neutrino effects can be described by scale-independent coefficients that are easy to implement in existing IC codes. To validate our approach, we constructed ICs based on a rigorous treatment of the scale-dependent neutrino response in 2LPT, obtaining agreement with our scheme to better than one part in $10^5$ up to $k=1\text{ Mpc}^{-1}$ in the power spectrum of the evolved CDM and baryon perturbations at late times.

Compared to these small differences, we find that the truncation error associated with using the first-order Zel'dovich approximation is much larger. For our fiducial model with $\sum m_\nu = 0.3$ eV and a starting redshift of $z_i=31$, the error is $4\%$ in the power spectrum and $7\%$ in the equilateral bispectrum around $k=0.5\text{ Mpc}^{-1}$ at $z=0$. Ratios of statistics from simulations with different neutrino masses can be calculated much more robustly, provided that the LPT order and starting redshift are the same. Nevertheless, even such ratios have a residual dependence on the ICs. For instance, Zel'dovich ICs introduce a mass-dependent error in the suppression of the power spectrum that grows with wavenumber $k$ and redshift $z$, peaking around the turn-over of the spoon. We also find that the starting time of the simulation has an impact on the suppression over a wide range of scales and redshifts. Simulations started at $z_i=127$ overestimate the suppression of the power spectrum on small scales, compared to later starts. While simulations can be started at higher redshifts to reduce truncation errors, this also increases the importance of particle resolution and relativistic effects. To minimize errors from initial conditions and particle resolution, simulations can be started at late times using higher-order ICs.

A major target of cosmological surveys is to measure the sum of neutrino masses. Assuming the minimum value allowed under the normal mass ordering, $\sum m_\nu = 0.06$ eV, cosmology could provide a $3\sigma$ detection and rule out the inverted mass ordering at $2\sigma$ by reaching a sensitivity of $0.02\text{ eV}$, which is in reach of future cosmic microwave background and large-scale structure experiments \citep{hamann12,abazajian15,brinckmann19,chudaykin19}. This corresponds to detecting $1\%$ effects on the matter power spectrum on $0.1\text{ Mpc}^{-1}<k<1\text{ Mpc}^{-1}$ scales. We should therefore aim for neutrino simulations with errors that are well below $1\%$ on these scales. While Zel'dovich ICs fall short of this mark, our findings suggest that 2LPT is sufficiently accurate for most applications. Higher-order statistics at high redshift seem to be the notable exception, which could be relevant for Lyman-$\alpha$ forest simulations.

The accuracy of neutrino simulations depends on many factors: the accuracy of the linear transfer functions and back-scaling procedure \citep{lesgourgues11b,zennaro17}, the implementation of neutrino perturbations (e.g. \citealt{bird18,elbers20}), neutrino initial conditions (Elbers, in prep.), and dark matter and baryon initial conditions (this paper). It has now been demonstrated that each of these factors can be controlled to within $1\%$. The remaining uncertainty is likely dominated by the choice of gravity solver. Achieving $1\%$ agreement between different $N$-body codes is non-trivial even in the absence of neutrinos \citep{schneider16,garrison19,grove21}. Fortunately, the accuracy of $N$-body codes should not in the first place be expected to deteriorate in the presence of neutrinos. In fact, the accuracy could even improve for particle-based implementations due to `spectator' effects (section \ref{sec:sims}). A systematic comparison of neutrino simulations with different codes and identical initial conditions could establish whether this is indeed the case. Such explorations would improve our ability to simulate nonlinear clustering in Universes with massive neutrinos, allowing us to meet the demands of the next generation of surveys.

\section*{Acknowledgements}

We thank Oliver Hahn, Cornelius Rampf, Matthieu Schaller, and the members of the Euclid Neutrino Method Comparison Project for useful discussions. We made extensive use of the \textsc{monofonIC} and \textsc{swift} codes and thank all contributors. WE is supported by the Durham Prize Scholarship in Astroparticle Physics. We acknowledge support from the European Research Council through ERC Advanced Investigator grant, DMIDAS [GA 786910] to CSF. BL is supported by the European Research Council (ERC) through ERC starting Grant No.~716532, and STFC Consolidated Grant (Nos.~ST/I00162X/1, ST/P000541/1). SP acknowledges partial support from the European Research Council under ERC Grant NuMass (FP7-IDEAS-ERC ERC-CG 617143), the European Unions Horizon 2020 research and innovation programme under grant No. 860881 -- HIDDeN -- H2020-MSCA-ITN-2019/H2020-MSCA-ITN-2019. This work used the DiRAC@Durham facility managed by the Institute for Computational Cosmology on behalf of the STFC DiRAC HPC Facility (www.dirac.ac.uk). The equipment was funded by BEIS capital funding via STFC capital grants ST/K00042X/1, ST/P002293/1 and ST/R002371/1, Durham University and STFC operations grant ST/R000832/1. DiRAC is part of the National e-Infrastructure.

\section*{Data availability}

All codes developed and used for this project are open source and available via \url{https://www.willemelbers.com/neutrino_ic_codes/}. Simulation products will be made available upon request to the corresponding author, subject to availability of storage.

\appendix

\section{Difference and sum equations}\label{sec:three_fluid_considerations}

As in (\ref{eq:euler_Dinf_euler}--\ref{eq:euler_Dinf_poisson}), the component fluid equations (\ref{eq:og_euler}--\ref{eq:og_cont}) can be rewritten using $D_\infty$ as time variable and $\boldsymbol{v}_\lambda=\boldsymbol{u}_\lambda/\partial_\tau D_\infty$ as velocity:
\begin{align}
	&\partial_{D_\infty}\boldsymbol{v}_\lambda + \boldsymbol{v}_\lambda\cdot\nabla_{\boldsymbol{x}}\boldsymbol{v}_\lambda = -\frac{3g_\infty}{2D_\infty}(\boldsymbol{v}_\lambda + \nabla_{\boldsymbol{x}}\varphi),  \label{eq:comp_euler_Dinf_euler}\\
	&\partial_{D_\infty}\delta_\lambda + \nabla_{\boldsymbol{x}}\cdot\left[(1+\delta_\lambda)\boldsymbol{v}_\lambda\right] = 0,  \label{eq:comp_euler_Dinf_continuity}
\end{align}

\noindent
for $\lambda\in\{\text{c},\text{b}\}$ with $\varphi = a\Phi/(B_0 D_\infty)$ and $g_\infty$ defined in \eqref{eq:g_inf_def}. The initial conditions at $D_\infty=0$ must be $\boldsymbol{v}_\text{c}=\boldsymbol{v}_\text{b}=-\nabla_{\boldsymbol{x}}\varphi$ for \eqref{eq:comp_euler_Dinf_euler} not to diverge. Taking the difference of \eqref{eq:comp_euler_Dinf_euler} for $\lambda=\text{b}$ and $\lambda=\text{c}$ gives
\begin{align}
    &\partial_{D_\infty}\boldsymbol{v}_\bc + \boldsymbol{v}_\text{b}\cdot\nabla_{\boldsymbol{x}}\boldsymbol{v}_\bc + \boldsymbol{v}_\bc\cdot\nabla_{\boldsymbol{x}}\boldsymbol{v}_\text{c}  = -\frac{3g_\infty}{2D_\infty}\boldsymbol{v}_\bc,
\end{align}

\noindent
where $\boldsymbol{v}_\bc=\boldsymbol{v}_\text{b}-\boldsymbol{v}_\text{c}$. Notice that the neutrino contribution contained in $\nabla_{\boldsymbol{x}}\varphi$ has dropped out. Consequently, we obtain results analogous to the $\Lambda$CDM case without massive neutrinos \citep{rampf21}. Expand $\boldsymbol{v}_\lambda = \sum_{m=1}^\infty \boldsymbol{v}^{(m)}_\lambda$ for $\lambda\in\{\text{c},\text{b}\}$ and $\boldsymbol{v}_\bc = \sum_{m=1}^\infty \boldsymbol{v}^{(m)}_\bc$. At first order, we find
\begin{align}
    &\partial_{D_\infty}\boldsymbol{v}^{(1)}_\bc = -\frac{3g_\infty}{2D_\infty}\boldsymbol{v}^{(1)}_\bc.
\end{align}

\noindent
Since $g_\infty$ is strictly positive (see Fig. \ref{fig:approx_validity_1}), the only non-decaying solution is $\boldsymbol{v}^{(1)}_\bc=0$. As $\boldsymbol{v}_\bc=0$ initially, this is the only solution. Suppose that $\boldsymbol{v}^{(m)}_\bc=0$ for $m=1,\dots,n-1$. Then also
\begin{align}
    &\partial_{D_\infty}\boldsymbol{v}^{(n)}_\bc = -\frac{3g_\infty}{2D_\infty}\boldsymbol{v}^{(n)}_\bc,
\end{align}

\noindent
with the only solution being $\boldsymbol{v}^{(n)}_\bc=0$. It follows that $\boldsymbol{v}_\bc=0$ at all orders. Using this result and taking the mass-weighted average of the component equations yields at all orders:
\begin{align}
	&\partial_{D_\infty}\boldsymbol{v}_\cb + \boldsymbol{v}_\cb\cdot\nabla_{\boldsymbol{x}}\boldsymbol{v}_\cb = -\frac{3g_\infty}{2D_\infty}(\boldsymbol{v}_\cb + \nabla_{\boldsymbol{x}}\varphi),\\
	&\partial_{D_\infty}\delta_\cb + \nabla_{\boldsymbol{x}}\cdot\left[(1+\delta_\cb)\boldsymbol{v}_\cb\right] = 0.
\end{align}

\noindent
Converting back to $\tau$-time gives (\ref{eq:euler1}--\ref{eq:euler2}). Letting $\delta_\bc = \delta_\text{b} - \delta_\text{c}$ and taking the difference of \eqref{eq:comp_euler_Dinf_continuity} for $\lambda=\text{b}$ and $\lambda=\text{c}$ also gives
\begin{align}
   &\partial_{D_\infty}\delta_\bc + \nabla_{\boldsymbol{x}}\cdot\left[\delta_\bc\boldsymbol{v}_\cb\right] = 0.
\end{align}

\noindent
Inserting $\delta_\bc = \sum_{m=1}^\infty \delta^{(m)}_\bc$, we find that $\delta^{(1)}_\bc=\text{constant}$ at first order, as in the case without neutrinos.

\section{Analytic solution}\label{sec:algebra}

We seek a solution to
\begin{align}
	\partial_\tau^2 D + aH\partial_\tau D = \frac{B_0}{a}D.  \label{eq:asymp_growing_tau_repeat}
\end{align}

\noindent
To express the solution as a function of the scale factor, $a(\tau)$, we switch time variables to $\log a$ and define the new velocity variable, $\boldsymbol{\tilde{u}}_\cb=\boldsymbol{u}_\cb/(aH)$. Eq.~\eqref{eq:asymp_growing_tau_repeat} is then written as
\begin{align}
	\frac{\mathrm{d}^2 D}{\mathrm{d}(\log a)^2} + \left[2 + \frac{\mathrm{d}\log H}{\mathrm{d}\log a}\right]\frac{\mathrm{d}D}{\mathrm{d}\log a} = \frac{B_0}{a^3 H^2}D  \label{eq:asymp_growing2}.
\end{align}

\noindent
The hypergeometric function $\geom(c,d,e,z)$ is a solution of the differential equation
\begin{align}
    z(1-z)\frac{\mathrm{d}^2 F}{\mathrm{d}z^2} + \left[e-(c+d+1)z\right]\frac{\mathrm{d}F}{\mathrm{d}z} - cd F = 0. \label{eq:geom_eq}
\end{align}

\noindent
Given the Ansatz $D(a)=a^p\sqrt{1+\Lambda a^3}F(z)$ with $z=-\Lambda a^3$ and $\Lambda=\Omega_\Lambda/\Omega_\text{m}$, we obtain after some algebra
\begin{align}
    \begin{split}
	&(1-z)\frac{\mathrm{d}^2F}{\mathrm{d}(\log a)^2} + \left[2(p+1)(1-z)-3z-\frac{3}{2}\right]\frac{\mathrm{d}F}{\mathrm{d}\log a} =\\
    &-\left[\left(p^2+\frac{p}{2}-\frac{3}{2}(1-f_\nu)\right)-\left(p^2+5p+\frac{21}{4}\right)z\right]F.
    \end{split} \label{eq:intermediate}
\end{align}

\noindent
To bring this in the form of \eqref{eq:geom_eq}, we require
\begin{align}
    p = \frac{1}{4}\left(\pm\sqrt{1+24(1-f_\nu)}-1\right),
\end{align}

\noindent
where the positive sign picks the growing solution. Using this in \eqref{eq:intermediate}, we obtain
\begin{align}
    \begin{split}
	&z(1-z)\frac{\mathrm{d}^2F}{\mathrm{d}z^2} + \frac{1}{3}\left[2p+\frac{7}{2}-(2p+8)z\right]\frac{\mathrm{d}F}{\mathrm{d}z} =\\
    &\frac{1}{9}\left[p^2+5p+\frac{21}{4}\right]F.
\end{split} \label{eq:intermediate2}
\end{align}

\noindent
Identifying constants in \eqref{eq:geom_eq} and \eqref{eq:intermediate2}, we derive the desired expression
\begin{align}
    D(a)=a^p\sqrt{1+\Lambda a^3}\geom\left(\frac{2p+7}{6}, \frac{2p+3}{6}, \frac{4p+7}{6},-\Lambda a^3\right),
\end{align}

\noindent
with $p=\sqrt{1+24(1-f_\nu)}/4-1/4$.

\section{Frame lagging}\label{sec:FL}

Let $S(\boldsymbol{x}) = \left(\delta_\cb*\alpha\right)(\boldsymbol{x})$. Since $S$ is itself first order, we have up to second order that
\begin{align}
	S(\boldsymbol{x}) = S(\boldsymbol{q}+\bm{\psi}) = S(\boldsymbol{q}) + \left.\frac{\partial S}{\partial q_i}\right\rvert_{\boldsymbol{q}}\psi_i(\boldsymbol{q}).
\end{align}

\noindent
Denoting the Fourier transform of $S(\boldsymbol{x})$ as $\mathcal{F} \left\{S(\boldsymbol{x})\right\}$, we find that
\begin{align}
	\mathcal{F}\left\{S(\boldsymbol{x})\right\} &= \mathcal{F}\left\{S(\boldsymbol{q})\right\} + \mathcal{F}\bigg\{\!\!\left.\frac{\partial S}{\partial q_i}\right\rvert_{\boldsymbol{q}}\!\bigg\}*\mathcal{F}\left\{\psi_i(\boldsymbol{q})\right\}.
\end{align}

\noindent
To be more explicit, we will denote the Fourier transform of $S(\boldsymbol{x})$ by $S^x(\boldsymbol{k})$ and the Fourier transform of $S(\boldsymbol{q})$ by $S^q(\boldsymbol{k})$. The above identity can then be written as
\begin{align}
	S^x(\boldsymbol{k}) = S^q(\boldsymbol{k}) + \int_{\boldsymbol{k}_1,\boldsymbol{k}_2}i\boldsymbol{k}_1^iS^q(\boldsymbol{k}_1)\psi_i(\boldsymbol{k}_2),
\end{align}

\noindent
where $\int_{\boldsymbol{k}_1,\boldsymbol{k}_2} = \int\mathrm{d}\boldsymbol{k}_1\mathrm{d}\boldsymbol{k}_2(2\pi)^{-6}\delta^{(3)}(\boldsymbol{k}_1+\boldsymbol{k}_2-\boldsymbol{k})$. Similarly,
\begin{align}
	\delta_\cb^q(\boldsymbol{k}) = \delta_\cb^x(\boldsymbol{k}) - \int_{\boldsymbol{k}_1,\boldsymbol{k}_2}i\boldsymbol{k}_1^i\delta_\cb^q(\boldsymbol{k}_1)\psi_i(\boldsymbol{k}_2).
\end{align}

\noindent
Combining the last two equations, we obtain
\begin{align}
	\alpha^x(\boldsymbol{k})\delta_\cb^x(\boldsymbol{k}) &= \alpha^q(\boldsymbol{k})\delta_\cb^x(\boldsymbol{k}) - F(\boldsymbol{k}),
\end{align}

\noindent
where we denote the so-called ``frame-lagging'' terms by
\begin{align}
	F(\boldsymbol{k}) &= \int_{\boldsymbol{k}_1,\boldsymbol{k}_2}i\boldsymbol{k}_1^i\left[\alpha^q(\boldsymbol{k}) - \alpha^q(\boldsymbol{k}_1)\right]\delta_\cb^q(\boldsymbol{k}_1)\psi_i(\boldsymbol{k}_2).
\end{align}

\noindent
Now, since $\delta_\cb^x=1/J-1$, we obtain the result used in section \ref{sec:full}:
\begin{align}
	[\delta_\cb*\alpha](\boldsymbol{x}) = [(1/J-1)*\alpha](\boldsymbol{q}) - F(\boldsymbol{q}).
\end{align}

\noindent
We now rewrite the second-order frame-lagging terms using the Monge-Amp\`ere equation, obtaining
\begin{align}
	F^{(2)}(\boldsymbol{k}) &= \int_{\boldsymbol{k}_1,\boldsymbol{k}_2}\left[\alpha(\boldsymbol{k}) - \alpha(\boldsymbol{k}_1)\right]\boldsymbol{k}_1^i\boldsymbol{k}_1^j\psi^{(1)}_i(\boldsymbol{k}_2)\psi^{(1)}_j(\boldsymbol{k}_1).
\end{align}

\section{Terms up to third order}\label{sec:3LPT_derivation}

We give explicit expressions up to third order. For $n=2$, both the cubic term on the right-hand side of \eqref{eq:approx_general1} and the quadratic term on the right-hand side of \eqref{eq:approx_general2} vanish. Hence, only the quadratic term in \eqref{eq:approx_general1} contributes. Using $\epsilon_{ijk}\epsilon_{ipq}=\delta_{jp}\delta_{kq} - \delta_{jq}\delta_{kp}$, we find
\begin{align}
    \nabla\cdot\bm{\psi}^{(2)} &= -\frac{3g_\infty}{4+3g_\infty}\frac{1}{2}\left[\psi_{i,i}^{(1)}\psi_{j,j}^{(1)} - \psi_{i,j}^{(1)}\psi_{i,j}^{(1)}\right].
\end{align}

\noindent
The corresponding $\Lambda$CDM coefficient ($3/7$) is found by setting $g_\infty=1$. Dividing these coefficients, one obtains the scale-independent factor $C_2 = 7g_\infty/(4+3g_\infty)$. For $n=3$, we obtain two pieces from \eqref{eq:approx_general1} and one piece from \eqref{eq:approx_general2}, giving $\bm{\psi}^{(3)} = \bm{\psi}^{(3a)} + \bm{\psi}^{(3b)} + \bm{\psi}^{(3c)}$. Using $\det A_{ij} = (1/6)\epsilon_{ijk}\epsilon_{pqr}A_{ip}A_{jq}A_{kr}$, we can write these as
\begin{align}
    \nabla\cdot\bm{\psi}^{(3a)} &= -\frac{g_\infty}{2+g_\infty}\det \psi^{(1)}_{i,j},\\
    \nabla\cdot\bm{\psi}^{(3b)} &= -\frac{4+6g_\infty}{6+3g_\infty}\frac{1}{2}\left[\psi_{i,i}^{(1)}\psi_{j,j}^{(2)} - \psi_{i,j}^{(1)}\psi_{i,j}^{(2)}\right],\\
    \nabla\times\bm{\psi}^{(3c)} &= -\frac{1}{3}\nabla\psi_i^{(2)}\times\nabla\psi_{i}^{(1)}.
\end{align}

\noindent
The corresponding $\Lambda$CDM terms are again found by setting $g_\infty=1$. Expressing these in terms of potentials (\ref{eq:approx_2lpt}-\ref{eq:approx_2lpt_3}) and dividing the corresponding coefficients, we obtain the form given in section \ref{sec:disp} in terms of $C_1,C_2,C_3$.

%%%%%%%%%%%%%%%%%%%%%%%%%%%%%%%%%%%%%%%%%%%%%%%%%%



\bibliographystyle{mnras}
\bibliography{main}

%%%%%%%%%%%%%%%%%%%%%%%%%%%%%%%%%%%%%%%%%%%%%%%%%%

%%%%%%%%%%%%%%%%% APPENDICES %%%%%%%%%%%%%%%%%%%%%

\appendix

% Don't change these lines
% \bsp	% typesetting comment
\label{lastpage}
\end{document}